\documentclass[12pt]{iopart}
\usepackage{graphicx}
\usepackage[labelformat=simple]{subcaption}
\usepackage[ruled,vlined]{algorithm2e}
\usepackage[hidelinks]{hyperref}

\begin{document}

\title{Pile-Up Mitigation using Attention}

\author{B Maier$^1$$\ddagger$, S M Narayanan$^5$\footnote[2]{These authors contributed equally to this work.}, G de Castro$^2$, M Goncharov$^3$, Ch Paus$^3$, M Schott$^4$}

\address{$^1$ CERN, Geneva, Switzerland}
\address{$^2$ California Institute of Technology, Pasadena, U.S.A.}
\address{$^3$ Massachusetts Institute of Technology, Cambridge, U.S.A.}
\address{$^4$ Johannes-Gutenberg Universit\"at Mainz, Mainz, Germany}
\address{$^5$ Currently at Flagship Pioneering}
\ead{benedikt.maier@cern.ch, sid.m.narayanan@gmail.com}

\vspace{10pt}
%\begin{indented}
%\item[]October 2021
%\end{indented}

\begin{abstract}
Particle production from secondary proton-proton collisions, commonly referred to as pile-up, impairs the sensitivity of both new physics searches and precision measurements at LHC experiments. We propose a novel algorithm, \textsc{Puma}, for modelling pile-up with the help of deep neural networks based on sparse transformers. These attention mechanisms were developed for natural language processing but have become popular in other applications. In a realistic detector simulation, our method outperforms classical benchmark algorithms for pile-up mitigation in key observables. It provides a perspective for mitigating the effects of pile-up in the high luminosity era of the LHC, where up to 200 proton-proton collisions are expected to occur simultaneously.
\end{abstract}

\vspace{2pc}
\noindent{\it Keywords}: LHC, HL-LHC, Pile-Up, Transformers, Machine Learning
%\maketitle

\section{Introduction}
\label{sec:intro}

The Large Hadron Collider (LHC) at CERN, Geneva, will remain the most powerful tool on Earth to produce and study heavy elementary particles, at least for this decade. Further maximizing the potential of its experiments, such as ATLAS and CMS, and thereby increasing the chances of discovering new physics in the coming LHC runs, is of paramount importance. These runs will be characterized by an ever-increasing instantaneous luminosity, i.e., by a larger average number of simultaneous proton-proton collisions. During the final High Luminosity phase of the LHC (HL-LHC) starting in 2027, this number is expected to reach 200, which is almost an order of magnitude more than what was seen during Run~2 (2016-2018).

This poses an enormous challenge to the experiments, because their reconstruction algorithms have to identify interesting signatures among a large number of signals coming from secondary collisions. Reconstructed objects falsely attributed to the primary collision are called ``pile-up'', and they can dramatically impair the sensitivity of an analysis. Especially in the Run~3 and High-Luminosity scenarios, removing pile-up contamination will become a primary objective. We present a method for identifying pile-up particles with the help of deep neural networks based on sparse transformers adapted from the field of natural language processing.

One example of a widely-used algorithm for rejecting pile-up particles is the charged-hadron subtraction as used in the CMS particle-flow (PF) algorithm~\cite{particleflow}. It removes charged particles whose tracks have not been assigned to the primary vertex. As more sophisticated algorithms, PUPPI~\cite{Bertolini2014} and SoftKiller~\cite{SoftKiller} aim at further identifying the pile-up component present among neutral particles.

First studies on the performance of pile-up mitigation using image recognition techniques~\cite{pumml} to identify pile-up contributions, and graph-based~\cite{vlimant,mikuni2020abcnet} methods have focused on a subset of particles in the event clustered into hadronic jets or did not consider detector resolution effects. In addition, the weights calculated for each particle using the PUPPI algorithm are used as input features in these networks. Here, we present for the first time a machine learning algorithm relying  only on raw reconstructed information that outperforms the classical benchmarks like PUPPI on event- and jet-level metrics, and in a realistic detector setting. This is a crucial step towards demonstrating the superiority of machine learning-based pile-up rejection on a global  event level at future detectors and scenarios like the HL-LHC.
% The point is not to show superiority of machine learning .... or is it?

The paper is organized as follows. Firstly, a description of the setup and of the datasets is provided that we use for training and for performing the comparisons between the different pile-up mitigation algorithms. We then give a detailed explanation of the implementation of \textsc{Puma}. Third, we introduce the key metrics to benchmark the performance of the algorithms and provide details on the algorithm training. Finally, we quantify the performance of \textsc{Puma}.

\section{Setup}
\label{sec:data}

\subsection{The \textsc{Delphes} Simulation Framework}

The goal of \textsc{Delphes}~\cite{delphes} is to allow the simulation of a multipurpose detector for phenomenological studies. The simulation includes a track propagation system embedded in a magnetic field, electromagnetic and hadron calorimeters, and a muon identification system. Physics objects that can be used for data analysis are then reconstructed from the simulated detector response. These include tracks and calorimeter deposits and high level objects such as isolated electrons, jets, taus, and missing energy. 

The \textsc{Delphes} framework allows for a fast simulation of an approximated detector response for typical LHC detectors using parameterized resolution and efficiency functions. The simulation includes a tracking system within a magnetic field, electromagnetic (ECAL) and hadronic calorimeters (HCAL) as well as a muon systems. High-level objects like isolated electrons, particle jets or missing transverse energies are reconstructed using low level observables such as tracks and energy deposits in the calorimeters. The calorimeter systems within \textsc{Delphes} is finely segmented in $\eta$ and $\phi$ and it is assumed that ECAL and HCAL have the same granularity, i.e. each ECAL cell has a corresponding cell in the HCAL. The geometrical center of each cell then defines the coordinate of the calorimeter energy deposit. 

The stable charged particles on generator level with a minimal transverse momentum (e.g. $p_\mathrm{T}>100$\,MeV) undergo the track reconstruction. The track reconstruction efficiency as well as the resolution and scale is parameterized vs. $p_\mathrm{T}$, $\eta$ and $\phi$. 

Particles on generator level that reach the calorimeter system leave the fractions $f_\mathrm{ECAL}$ and $f_\mathrm{HCAL}$ in the electromagnetic and hadronic calorimeter cells, respectively. The relevant cells can then be group together in one calorimeter tower. When several particles reach the same cells the total energy of one tower is simple the sum over all particles, that leave energies either in the ECAL or HCAL. \textsc{Delphes} assumes that all electrons and photons leave their full energy on the ECAL system, i.e. $f_\mathrm{ECAL}=1.0$ and $f_\mathrm{HCAL}=0.0$. Muons and neutrinos are assumed to leave no energy at all in the calorimeter system. All stable hadrons are treated with $f_\mathrm{ECAL}=0.0$ and $f_\mathrm{HCAL}=1.0$, with the exception of Kaons and $\Lambda$, where the values $f_\mathrm{ECAL}=0.3$ and $f_\mathrm{HCAL}=0.7$ are used.

The resolutions of the electromagnetic and the hadronic calorimeters are independently parameterized in dependence of the particle kinematics, using a stochastic, a noise and a constant as parameters. 

The CMS collaboration was one of the first that implemented a PF algorithm for the reconstruction and measurement of finale state objects, in order to maximize the use of sub-detector measurements for the event reconstruction. Since the full PF approach is rather complex, a simplified version is implemented within the \textsc{Delphes} framework, which results in two types of object collections: PF tracks and PF towers. Each tower is described by the total energy deposited in the electromagnetic and the hadronic calorimeter, $E_\mathrm{ECAL}$ and $E_\mathrm{HCAL}$, respectively. In addition, the total energy deposit originating from charged particles in a given tower is stored in the variables $E_\mathrm{ECAL,trk}$ and $E_\mathrm{HCAL,trk}$. It is important to note in this context, that \textsc{Delphes} assumes that the momentum of charged particles is always measured best by the tracking system. This allows to define the energy flow of a tower by

\[E_\mathrm{Tower}^\mathrm{eflow} = \max(0, \Delta_\mathrm{ECAL})+\max(0, \Delta_\mathrm{HCAL}),\]

where $\Delta_\mathrm{ECAL}=E_\mathrm{ECAL}-E_\mathrm{ECAL,trk}$ and $\Delta_\mathrm{HCAL}=E_\mathrm{HCAL}-E_\mathrm{HCAL,trk}$. The PF algorithms then create a PF track for each reconstructed track and a PF tower with the energy $E_\mathrm{Tower}^\mathrm{eflow}$ if $E_\mathrm{Tower}^\mathrm{eflow}>0$. 

Particle-flow tracks describe therefore all charged particles with a good resolution. Particle-flow towers on the other hand, contain the energy information of neutral particles and charged particles that have not been reconstructed by the tracking system. In addition, also energy deposits due to the positive smearing of the calorimeters are taken into account. A detailed description of the algorithms can be found in Reference~\cite{delphes}, where also several validation studies are shown.

\subsection{Collision datasets}

To emulate the HL-LHC data taking situation, we generate Monte-Carlo simulations of LHC proton-proton collisions at $\sqrt{\mathrm{s}}=\mathrm{14}$\,TeV using Pythia version 8.244~\cite{pythia8}. Pythia is used for both matrix-element generation and parton showering and hadronization, employing the parton shower tune 4C~\cite{Corke_2011}, which also provides a description of effects from multi-parton interaction and the underlying event. The stable generator-level particles are subsequently passed through a model of the CMS detector built with Delphes version 3.4.3pre01, retaining the correspondence between reconstructed PF objects and the incident truth particles. The layout of the detector roughly corresponds to the Phase-II upgrade conditions at CMS, including novel, high-granularity forward detector components. The processes simulated are dileptonic $\mathrm{t}\bar{\mathrm{t}}$ production, $\mathrm{Z}(\to\nu\bar{\nu})$+jets production, vector boson fusion (VBF) production of a Higgs boson with subsequent $\mathrm{H}\to \mathrm{c}\bar{\mathrm{c}}$ or $\mathrm{H}\to \mathrm{dark~matter}$ decays, and soft-QCD production. For each of the studied SM processes, 200 thousand events have been generated for training and evaluation. On average, 140 soft-QCD events are mixed with the hard scattering event, sampled from a total of 50 million soft-QCD events. In the following, ``PF candidate'' and ``particle'' will be used interchangeably to refer to a PF object.

Very few simplifying assumptions are made in the simulation and reconstruction of the events: for stable, charged particles (electrons, muons, charged hadrons), the vertex assignment is assumed to be perfect. No underlying event is simulated. We note that apart from these assumptions, this study is based on a fast but sophisticated detector simulation that realistically models particle reconstruction efficiencies and smearing of track momenta and calorimeter tower energies.

\subsection{Particle and event definitions}

Each PF candidate is represented by a set of features: its Lorentz four-vector, the particle species, the electric charge, the corresponding vertex (if available), and local cluster information (see Section~\ref{sec:architecture:cluster} for the method to obtain event sub-clusters). An event is represented as a set of PF candidates, as well as the number of reconstructed vertices. For each particle, we also compute the target quantity to be learned:
\begin{equation}
    y = \frac{E_\mathrm{LV}^\mathrm{gen}}{E_\mathrm{LV}^\mathrm{gen}+E_\mathrm{PU}^\mathrm{gen}} \ ,
\end{equation}
where $E_\mathrm{LV}^\mathrm{gen}$ is the summed energy of the incident generated particles from the leading vertex (LV), i.e., from the hard interaction, associated with the PF candidate. Similarly, $E_\mathrm{PU}^\mathrm{gen}$ is the summed energy of associated incident generated particles stemming from pile-up interactions. The quantity $y$ is referred to as hard energy fraction in the following.

\section{Modeling Pile-up with Sparse Transformers}
\label{sec:architecture}

We formulate the problem of pile-up identification as one of regressing the hard energy fraction of each particle. That is, we define a model $g(\cdot;\Theta)$ to minimize the loss $\mathcal{L}(\cdot; \Theta)$:
\begin{equation}
    \mathcal{L}\left(\{\mathbf{x}_i\}, \{y_i\}; \Theta\right) = \sum_j ||g(\{\mathbf{x}_i\};\Theta)_j - y_j||_2^2 \ ,
    \label{eq:loss_unweighted}
\end{equation}
where $\mathbf{x}_i$ is the feature vector of particle $i$ in a given event, $y_i$ is the hard energy fraction of particle $i$, $\Theta$ denotes free parameters of the model $g$, and $g(\cdots)_j$ is the prediction of the model for particle $j$. Note that this prediction is conditioned on all particles in the event $\{\mathbf{x}_i\}$, but the loss is computed independently for each particle.

This energy fraction regression task is distinct from previous ML approaches to the pile-up problem~\cite{vlimant,mikuni2020abcnet}, which treat the problem as one of classification: a particle is uniquely identified as arising from a hard or pile-up vertex. This definition is no longer valid when considering the realistic scenario of a detector with finite spatial and energy resolutions. A measured PF candidate is frequently the result of several particles depositing energy in the same detector component. Accordingly, we train our models to decompose each detected particle into co-linear LV and PU components.

We parameterize $g$ as a deep neural network \textsc{Puma}: Pile-Up Mitigation using Attention. Attention was first introduced in 2014~\cite{firstatt} for neural machine translation of natural languages, and further developed as self-attending Transformer networks~\cite{vaswani2017} for a multitude of natural language tasks. 

The dynamics of particle decay and hadronization at the LHC share some conceptual similarities with natural language problems: much information of particle dynamics can be described in a local picture (in phase space), with some information requiring higher-order global abstraction (e.g. jet hadronization). This is analogous to a natural language sentence, where most words are closely coupled to nearby words, but some long-range dependencies do arise. Therefore, we hypothesize that a similar model parameterization may work in both scenarios. 

At the core of \textsc{Puma} lie several transformer layers~\cite{vaswani2017}. The full model can be written as:

\begin{equation}
\eqalign{
    h_\mathrm{embed.} = \mathrm{MLP}_\mathrm{embed.}(X) \nonumber \cr
    h_\mathrm{enc.} = \textsc{Transformer}(h_\mathrm{embed.}) \cr  
    \hat{Y} = \mathrm{MLP}_\mathrm{dec.}(h_\mathrm{enc.}) \ ,}
\end{equation}

where $X = [\mathbf{x}_1,\dots,\mathbf{x}_N]^T$ is the particle feature matrix, $\mathrm{MLP}$ are multi-layer perceptrons, and \textsc{Transformer} is a stack of Transformer layers.

In defining a model that uses full self-attention, a difficulty arises: the time and memory complexity of self-attention scales quadratically with the cardinality of the input set. In a typical particle collision with 140 pile-up interactions, the number of particles, $N$, can reach ten thousand. This makes training the model with reasonable batch sizes, even on large GPUs, prohibitively difficult.

To reduce the memory consumption of the transformer, we find a way to sparsify the self-attention mechanism. While many variants of sparse attention exist~\cite{sparse1,sparse2,sparse3}, we use the Longformer~\cite{beltagy2020} implementation. Essentially, we construct a nearest-neighbors graph among the particles, such that the non-zero elements of the adjacency matrix are a subset of a banded-diagonal matrix. By limiting the size of this band, $w$, we ensure that the attention complexity scales as $\mathcal{O}(Nw) \ll \mathcal{O}(N^2)$.

\subsection{Hierarchical particle clustering and sparsification}
\label{sec:architecture:cluster}

First, we cluster particles in a hierarchical fashion on the surface of a cylinder parameterized by $(\sin(\phi), \cos(\phi), \eta)$. The result of $\textsc{IterativeCluster}(\{\mathbf{x_i}\}, w_\mathrm{clust}, k)$ (Algorithm~\ref{algo:cluster}) is a set of clusters of particles, of size $w_\mathrm{clust}$ or smaller.

\begin{algorithm}[H]
    \SetAlgoVlined 
    \SetKwProg{Fn}{Function}{ :}{}
    \Fn{\textsc{IterativeCluster}($C$, $w_\mathrm{clust}$, $k$) $\rightarrow$ clusters}{
        \eIf{$|C| < w_\mathrm{clust}$} {
            return $\{C\}$ \; 
        }{
            $\{C_1,\dots,C_k\} \leftarrow \textsc{KMeans}(k, C)$ \;
            $FC \leftarrow \{\}$ \; 
            \For{$i=1,\dots,k$}{
                $FC \leftarrow FC ~\bigcup ~ \textsc{IterativeCluster}(C_i, w_\mathrm{clust}, k)$ \;
            }
            return $FC$ \;
        }
    }
    \caption{\label{algo:cluster}
      Hierarchical particle clustering algorithm: $w_\mathrm{clust}$ is the maximum cluster size, and $k$ is the number of clusters at each iteration. \textsc{KMeans} is a $k$-means clustering function that computes clusters on a cylindrical surface.}
\end{algorithm}

\begin{figure}
  \centering
  \begin{subfigure}{0.475\textwidth}
    \includegraphics[width=\textwidth]{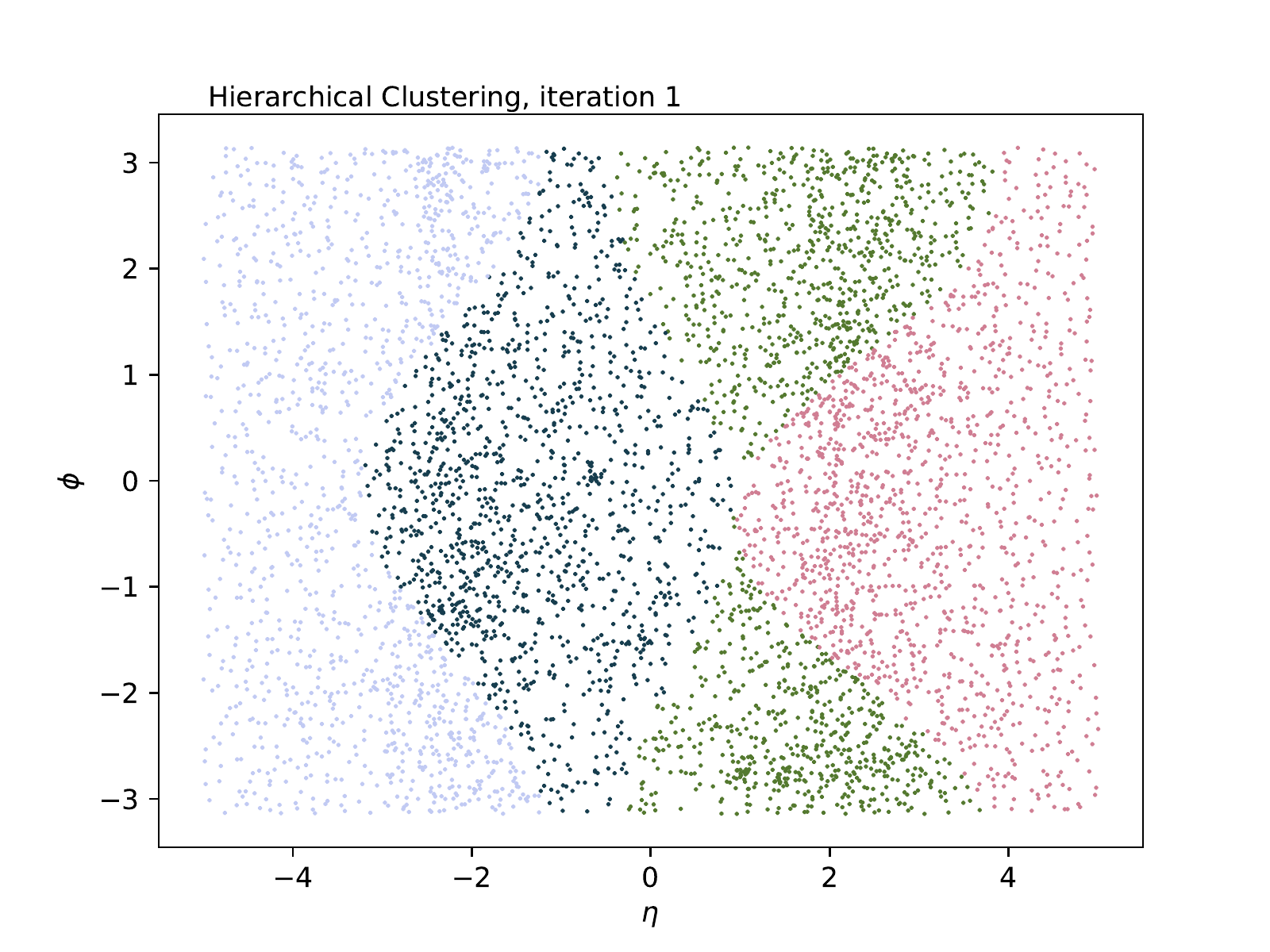}
    \caption{}
    \label{fig:cluster:1}
  \end{subfigure}
  \begin{subfigure}{0.475\textwidth}
    \includegraphics[width=\textwidth]{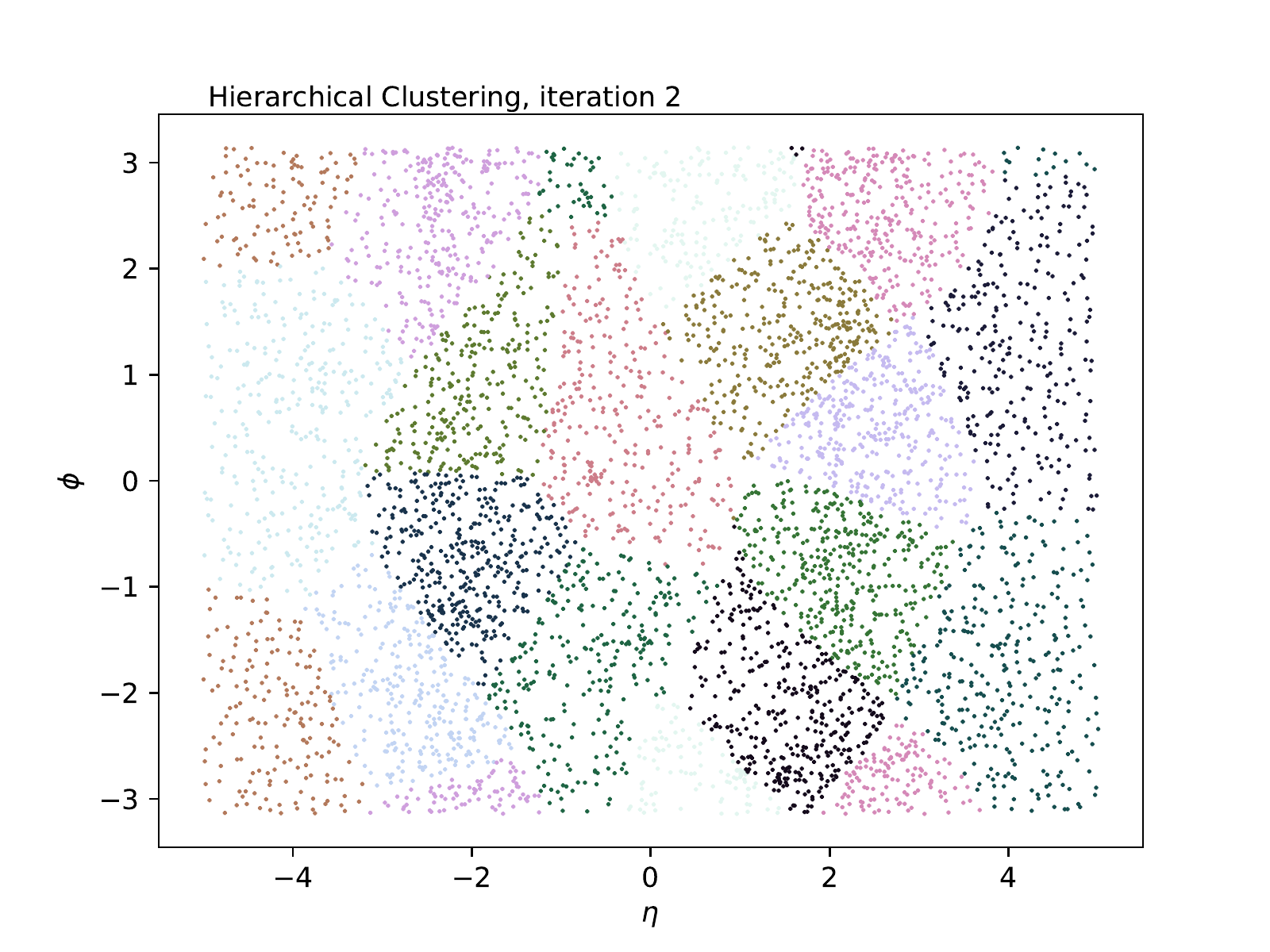}
    \caption{}
    \label{fig:cluster:2}
  \end{subfigure}
  \begin{subfigure}{0.475\textwidth}
      \includegraphics[width=\textwidth]{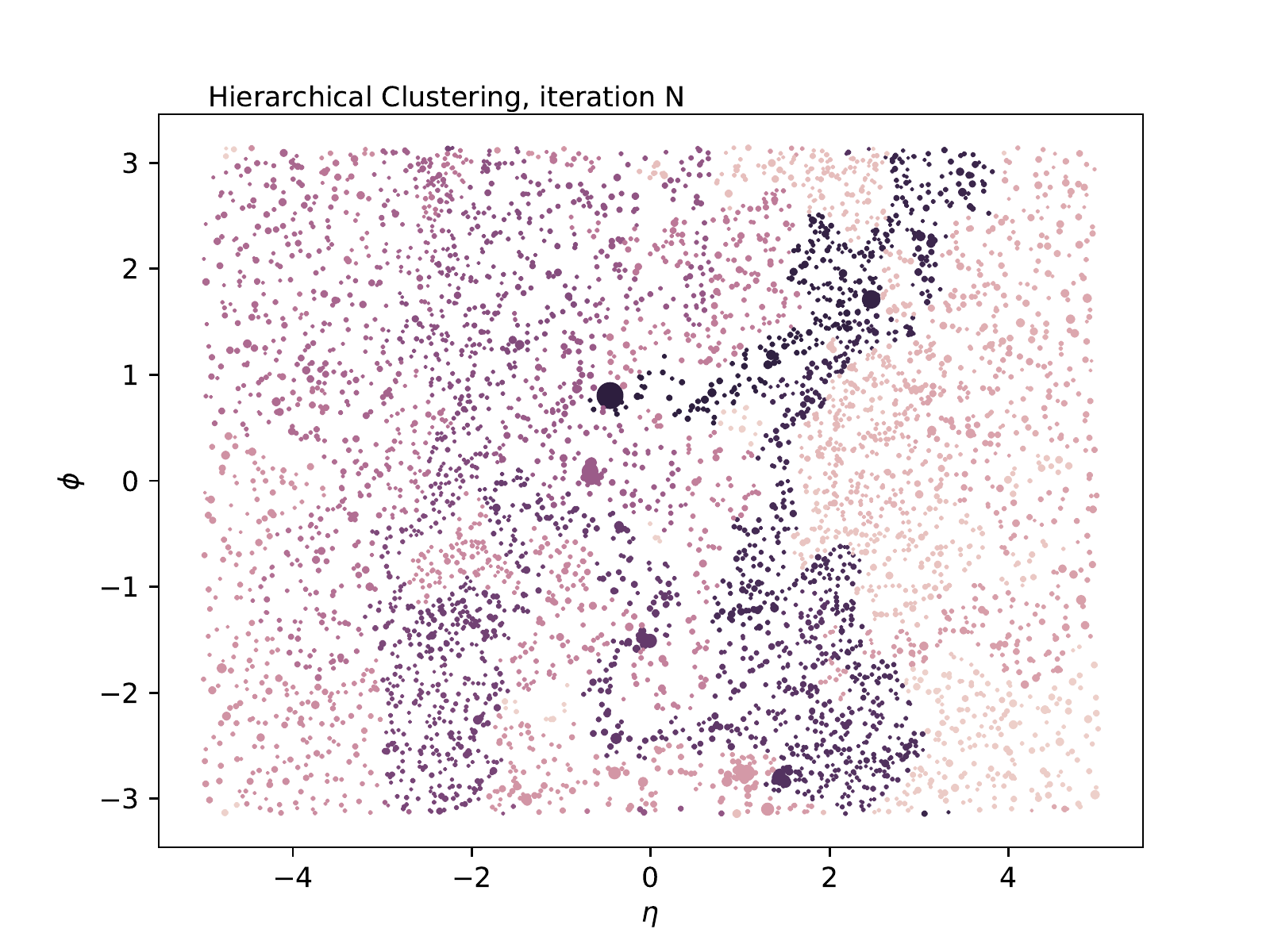}
      \caption{}
      \label{fig:cluster:N}
  \end{subfigure}
  \caption{Particle assignment to clusters (indicated by color) for a certain event after 1 iteration, after 2 iterations, and after convergence. In all three figures, $k=4$ and $w_\mathrm{clust}=10$. In Fig.~\ref{fig:cluster:N}, clusters are colored based on their ordering (from dark to light), and particle marker areas are proportional to particle $p_\mathrm{T}$.}
  \label{fig:cluster}
\end{figure}

Once the clusters $\{C_1,\dots,C_{n_c}\}$ have been computed, we define a unique ordering. For each cluster $C_i$, we compute three quantities:
\begin{equation}
    p_\mathrm{T}(C) = \sum_{j \in C} p_\mathrm{T}^j, \quad 
    \eta(C) = \sum_{j\in C} \frac{p_\mathrm{T}^j\cdot\eta^j}{p_\mathrm{T}(C)}, \quad 
    \phi(C) = \sum_{j\in C} \frac{p_\mathrm{T}^j\cdot\phi^j}{p_\mathrm{T}(C)}
\end{equation}
These represent, respectively, the total transverse momentum of the cluster, the $p_\mathrm{T}$-weighted pseudorapidity of the cluster, and the $p_\mathrm{T}$-weighted azimuthal angle of the cluster. We define a permutation of the clusters $\pi$. It is initialized as:
\begin{equation}
    \pi(0) = \mathrm{argmax}_i p_\mathrm{T}(C_i)
\end{equation}
Then, for $a>0$:
\begin{equation}
    \pi(a) = \mathrm{argmin}_i \left\{ \Delta R(C_i, C_{\pi(a-1)}) ~\Big|~ i \neq \pi(b) ~ \forall~ b < a \right\} \ ,
\end{equation}
where $\Delta R(x, y) = \sqrt{(\eta_x - \eta_y)^2 + (\phi_x - \phi_y)^2}$ is the $L_2$ metric in $(\eta,\phi)$ space. The results of this hierarchical clustering and ordering are shown for an example event in Fig.~\ref{fig:cluster}.

\begin{figure}
  \centering
  \begin{subfigure}{0.475\textwidth} \includegraphics[width=\textwidth]{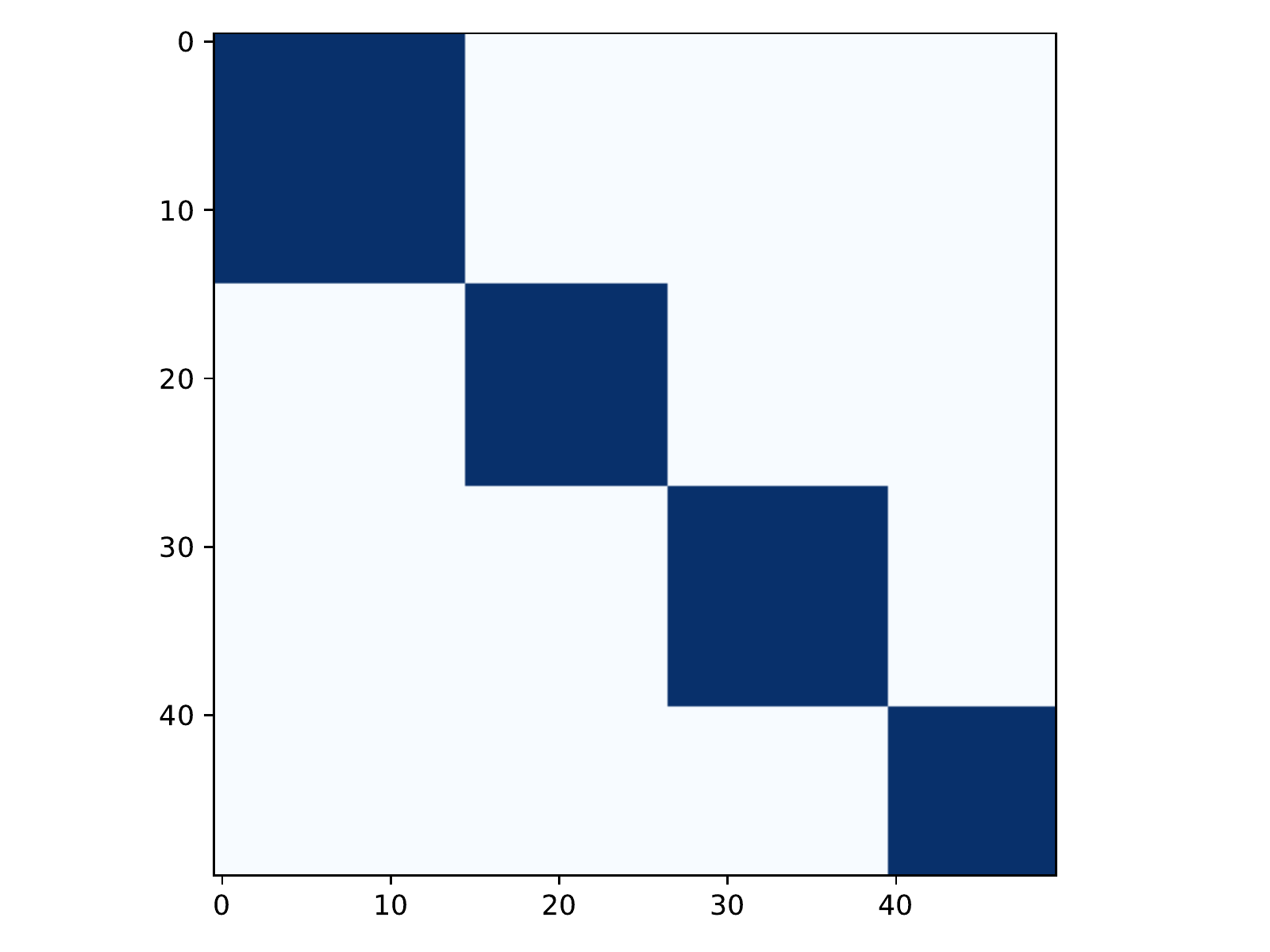} \caption{} \label{fig:adj:coarse} \end{subfigure}
  \begin{subfigure}{0.475\textwidth} \includegraphics[width=\textwidth]{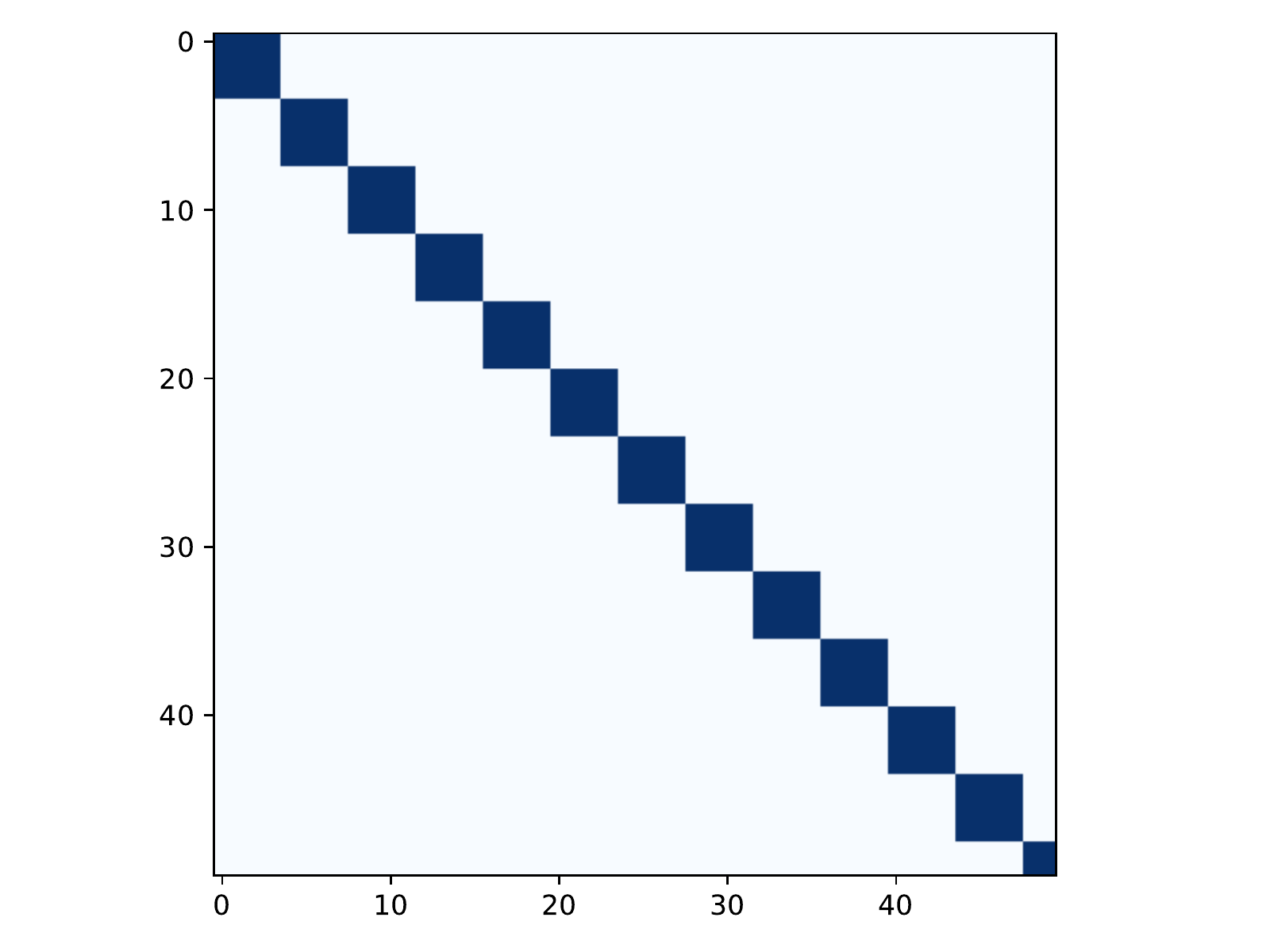} \caption{} \label{fig:adj:fine} \end{subfigure}
  \begin{subfigure}{0.475\textwidth} \includegraphics[width=\textwidth]{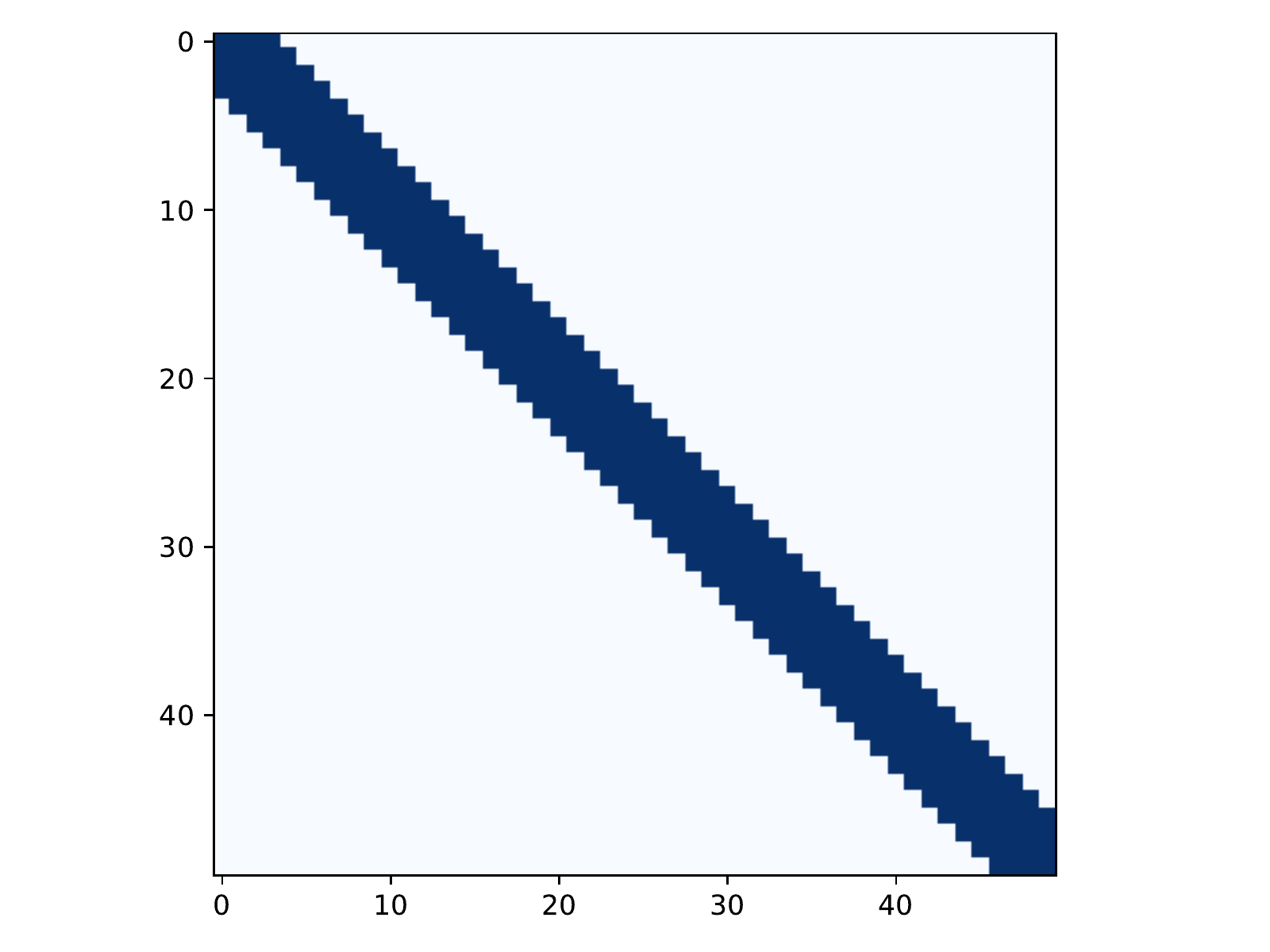} \caption{} \label{fig:adj:tight} \end{subfigure}
  \begin{subfigure}{0.475\textwidth} \includegraphics[width=\textwidth]{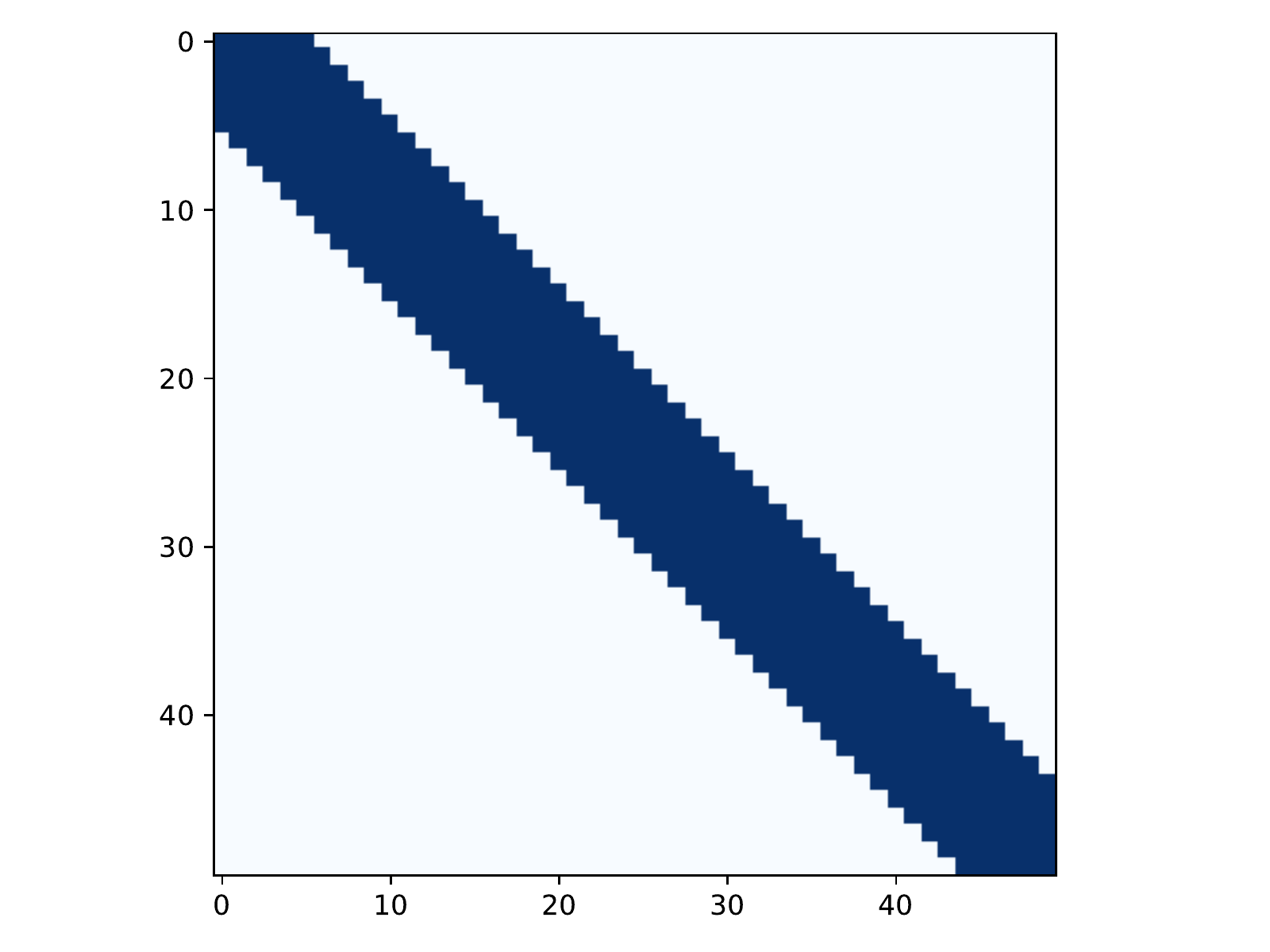} \caption{} \label{fig:adj:loose} \end{subfigure}
  \caption{Figs.~\ref{fig:adj:coarse}-\ref{fig:adj:fine}: possible adjacency matrices for a 50-particle event, with $w_\mathrm{clust}=15$ and $5$, respectively. 
  Fig.~\ref{fig:adj:tight}: the minimum banded-diagonal matrix required to cover the $w_\mathrm{clust}=5$ adjacency matrix. 
  Fig.~\ref{fig:adj:loose}: a slightly wider banded-diagonal matrix that allows for longer-range attention across clusters.}
  \label{fig:adj}
\end{figure}

The ordering of the clusters give a natural ordering of the particles, in which the particles are first ordered according to the cluster they belong to, and within each cluster, by decreasing $p_\mathrm{T}$.
The particles can then be thought of as a graph with $n_c$ complete, mutually-disconnected subgraphs. 
Each particle is a vertex, and two particles are connected by an edge if they belong two the same cluster. As a consequence of our particle ordering, the adjacency matrix $A_\mathrm{clust}$ is block-diagonal, as illustrated in Fig.~\ref{fig:adj}. Furthermore, our cluster ordering means each block is adjacent to blocks arising from clusters that are proximal in $(\eta,\phi)$-space. With $\mathrm{<}n_\mathrm{PU}\mathrm{>}=140$, the number of reconstructed particles per event averages at about 6000, with some events having up to 9000 particles. Therefore, the sequence of particles is zero-padded up to a length of 9000 if fewer particles are reconstructed.

\subsection{Banded attention}
\label{subsec:band}

The clustering allows us to avoid attending over particles that are far away from the query particle.
However, if we were to only allow the attention mechanism to consider keys adjacent to the query in $A_\mathrm{clust}$, the model could not consider relationships between clusters. Instead, we use a banded-diagonal matrix $A$ with bandwidth $w > w_\mathrm{clust}$. Larger values of $w$ allow stronger cross-cluster attention within a Transformer layer. Figure~\ref{fig:adj} illustrates the relationship between $w_\mathrm{clust},A_\mathrm{clust}$ and $w,A$. In practice, we find that values of $w$ two to four times larger than $w_\mathrm{clust}$ allow for sufficient information flow between clusters. Attention band widths smaller than $w=100$ allows us to leverage the efficient attention kernels provided by the authors of~\cite{beltagy2020}. For widths larger than 100, we encounter memory limitations on our hardware. However, by stacking Transformer layers, we are effectively still able to increase the receptive field, as the transformed feature vectors after layer $n$ carry information from particles $n \times w$ away from the query particle.

\textsc{Puma} is implemented in Python using PyTorch~\cite{pytorch} and uses elements of the Transformers~\cite{huggingface} library. The data processing, including \textsc{IterativeCluster}, is implemented in C++ using Delphes and ROOT~\cite{root}.

\section{Model Training and Evaluation}

\begin{table}
\caption{Variables used as input to the network. While they are attributes of each particle, they resemble three categories of different locality: variables corresponding to properties of the individual particle ($p_\mathrm{T}$, $\eta$, $\phi$, $E$, particle ID, vertex ID); variables characterizing the cluster the particle is in (cluster ID, cluster $R$, cluster $p_\mathrm{T}$); an event-wide variable ($N_\mathrm{PV}$).}
\begin{center}
\begin{footnotesize}
\begin{tabular}{ r  l }
  \hline
    \hline
  $p_\mathrm{T}$ & particle transverse momentum \\
  $\eta$ & particle pseudorapidity  \\
  $\phi$ & particle azimuthal angle \\
  $E$ & particle energy \\
  particle ID & particle ID \\
  vertex ID & vertex of particle, -1 if neutral \\
    \hline
  cluster ID & index of cluster containing particle \\
  cluster $\Delta R$ & max. pairwise distance $\Delta R = \sqrt{\Delta\eta^{2}+\Delta\phi^{2}}$ between particles found in cluster \\
  cluster $p_\mathrm{T}^\mathrm{ch}$ & scalar sum of $p_\mathrm{T}$'s of all charged LV particles in cluster containing the particle \\
    cluster $p_\mathrm{T}^\mathrm{neut}$ & scalar sum of $p_\mathrm{T}$'s of all neutral particles in cluster containing the particle \\
    \hline
  $N_\mathrm{PV}$ & number of reconstructed vertices in the event\\
  \hline
    \hline
\end{tabular}
\label{tab:inputvars}
\end{footnotesize}
\end{center}
\end{table}

\subsection{Training procedure}

The standard MSE loss (Eq.~\ref{eq:loss_unweighted}) puts all particles on equal footing. 
However, in a high pile-up event, we expect the vast majority of particles to be of extremely low-momentum, while the event dynamics are dominated by a handful of high-momentum particles. 
To bias \textsc{Puma} towards more accurate modeling of high-momentum particles, we modify the loss:
\begin{equation}
\eqalign{
    \mathcal{L}\left(\{\mathbf{x}_i\}, \{y_i\}; \Theta\right) = \sum_j f_j ||g(\{\mathbf{x}_i\};\Theta)_j - y_j||_2^2 \cr
    f_j = \frac{p_{\mathrm{T},j}^\alpha}{\max_k p_{\mathrm{T},k}^\alpha}\ ,}
    \label{eq:loss_weighted} 
\end{equation}
where $k$ is the index over the particles contained in the leading cluster and $\alpha \geq 0$ is a tunable hyperparameter. We choose $\alpha=2$ for our studies.

As vertex identification for charged particles is essentially perfect, \textsc{Puma} only needs to optimize the loss function in Eq.~\ref{eq:loss_weighted} for neutral particles. This is equivalent to setting $f_j=0$ for charged particles. In practice, we find doing so does not improve performance on neutral particle energy regression, nor other downstream metrics. In what follows, the loss is computed over all particles.

The loss function in Eq.~\ref{eq:loss_weighted} is minimized stochastically using the ~\textsc{Adam} optimizer~\cite{adam2014}. We employ batch sizes between 512 and 8192, distributed across four Nvidia Tesla V100 GPUs. To achieve the higher end of this range, we accumulate gradients over a maximum of 8 iterations before applying weight updates. The learning rate is initialized to $10^{-3}$ and follows a cyclical schedule~\cite{smith2017}, with a period of 4 epochs and decay factor of 0.97. All models are trained to convergence, which typically occurs after $\mathcal{O}(10^5)$ steps. %When training the network on a sample of 200\,000 dileptonic $t\bar{t}$ events, this means the network sees the same simulated event about ? times in the training phase.

\subsection{Evaluation metrics}
\label{subsec:metrics}

In addition to the key metric of the weighted mean square error, we evaluate the performance of pile-up mitigation techniques using three other metrics: the transverse momentum imbalance, ($p_\mathrm{T}^\mathrm{miss}$); the leading jet $p_\mathrm{T}$; and the RMS of the transverse component of the hadronic recoil vector.

The vector sum of the $p_\mathrm{T}$ of all produced particles must be zero because the initial state of a hadron collision has no net momentum in the transverse plane. We calculate this as variable as:
\begin{equation}
    \vec{p_\mathrm{T}}^\mathrm{miss} = -\sum_{i\in\mathrm{ particles}} \vec{p_{\mathrm{T}}}_{,i}
\end{equation}
In events with undetectable particles (e.g. neutrinos), we expect to find $p_\mathrm{T}^\mathrm{miss} \equiv |\vec{p_\mathrm{T}}^\mathrm{miss}| > 0$. Many crucial Standard Model measurements, e.g. of the W mass, involve neutrinos and rely on an optimal pile-up rejection, and several beyond-SM models predict other undetectable particles (e.g. dark matter, stable SUSY particles, gravitons). Pile-up interactions are isotropically distributed in $\phi$ and have minimal $p_\mathrm{T}^\mathrm{miss}$ which causes an increase of the variance of reconstructed $p_\mathrm{T}^\mathrm{miss}$. Therefore, it is of great importance to assess the impact of pile-up, and pile-up mitigation techniques, on the missing transverse momentum. We define three per-event metrics from the momentum imbalance:
\begin{equation}
    \hat{p}_x^\mathrm{miss} - p_x^\mathrm{miss}, \quad \hat{p}_y^\mathrm{miss} - p_y^\mathrm{miss}, \quad 
    \hat{p}_\mathrm{T}^\mathrm{miss} - p_\mathrm{T}^\mathrm{miss} \, ,
\end{equation}
where $p$ refers to the true momentum imbalance of generated particles prior to detector effects and $\hat{p}$ refers to the estimated momentum imbalance of reconstructed particles after detector effects, PF reconstruction, and pile-up mitigation.

As the LHC is a hadron collider, many events produce jets, i.e., bundles of collimated hadrons. In some cases, the jet is incidental to the process being studied (e.g. production of a Z boson at high $p_\mathrm{T}$), while in other cases it is a fundamental signature of the process (e.g. VBF production of a Higgs boson). Pile-up interactions typically produce soft jets. However, even if the hard interaction has produced several hard jets, particles from the soft pile-up jets can affect the reconstruction of the hardest jets. Therefore, we also consider misreconstruction in the mass of the dijet system in the VBF Higgs production mode as another metric:
\begin{equation}
    \hat{m}_{jj}^\mathrm{VBF} - m_{jj}^\mathrm{VBF}
\end{equation}

We do not expect any of these metrics to reach the ideal value of zero, even with perfect pile-up regression, because $\hat{p}$ includes the effects of detector resolution and PF reconstruction. Some previous work has neglected these effects, probably to isolate the impact of pile-up. %Finite detector resolution and PF reconstruction fundamentally impact pile-up identification (see Sec.~\ref{sec:architecture}). 

Finally, the measured hadronic recoil, $\vec{U}$, in a Z+jets event can be decomposed in a parallel, $U_{||}$, and a perpendicular, $U_{\perp}$, component with respect to the true vector boson transverse momentum, $p_\mathrm{T}^\mathrm{Z}$, where an ideal measurement yields $p_\mathrm{T}^\mathrm{Z}+U_{||}=0$ and $U_{\perp}=0$. The RMS error of the $U_{\perp}$ distribution is typically taken as a measure of the hadronic recoil resolution. However, this definition has the disadvantage that any rescaling of the measured hadronic recoil components by some factor $\alpha$ also changes the resolution.

\begin{equation} \label{eq:ComFac}
U'_{\perp} = \alpha \cdot U_{\perp}, \hspace{1cm} U'_{||} = \alpha \cdot U_{||}
\end{equation}

The width of the rescaled $U'_{\perp}$ distribution will be smaller for $\alpha<1$, however no real gain in a physics measurement has been achieved, because an additional bias in $p_\mathrm{T}^\mathrm{Z}+U_{||}$ is introduced and the sensitivity of $U_{||}$ on $p_\mathrm{T}^\mathrm{Z}$ decreases. In order to ensure a fair comparison between the different methods, the factors $\alpha_i$ in Eq.~\ref{eq:ComFac} for each bin of $p_\mathrm{T}^\mathrm{Z}$ have been chosen such that the average bias $<p_\mathrm{T}^\mathrm{Z}+U_{||}>$ in a given bin of $p_\mathrm{T}^\mathrm{Z}$ is the same for all methods. The figure of merit is then the RMS error of the resulting $U'_{\perp}$ distribution.

We define our gold standards as the event descriptions achievable assuming perfect pile-up regression but imperfect particle reconstruction.
That is, we consider four scenarios:
\begin{itemize}
    %\item Gold standard: scale each particle's 4-vector by the true hard energy fraction $y$.
    \item \textsc{CHS}: charged-hadron subtraction: only consider charged particles if they are associated with the primary vertex; leave neutral particles unscaled  
    \item Puppi: scale each particle's 4-vector by the pile-up likelihood $w_\mathrm{puppi}$
    \item \textsc{Puma}: scale each particle's 4-vector by the estimated hard energy fraction $\hat{y}$
    \item Gold standard: compare to a sample generated with $n_\mathrm{PU}=0$
\end{itemize}

\section{Results}\label{sec:results}

\subsection{Model hyperparameters}

In this section we describe the final choice of model hyperparameters selected. The cluster size used in \textsc{IterativeCluster} is set to $w_\mathrm{clust}=10$. The remainder of the hyperparameters are described in Tab.~\ref{tab:hps}. The model has 127074 trainable parameters. 

\begin{table}
    \caption{Our choice of \textsc{Puma} hyperparameters.}
    \centering
    \begin{tabular}{lc}
     \hline  \hline
        Parameter & Value  \\ \hline
        Embedding size & 64 \\ 
        Hidden layer size & 64 \\ 
        Number of attention heads & 4 \\ 
        Attention band width & 15 \\ 
        Number of Transformers & 12 \\ 
         \hline  \hline
    \end{tabular}
    \label{tab:hps}
\end{table}

\begin{figure}
  \centering
  \includegraphics[width=0.475\textwidth]{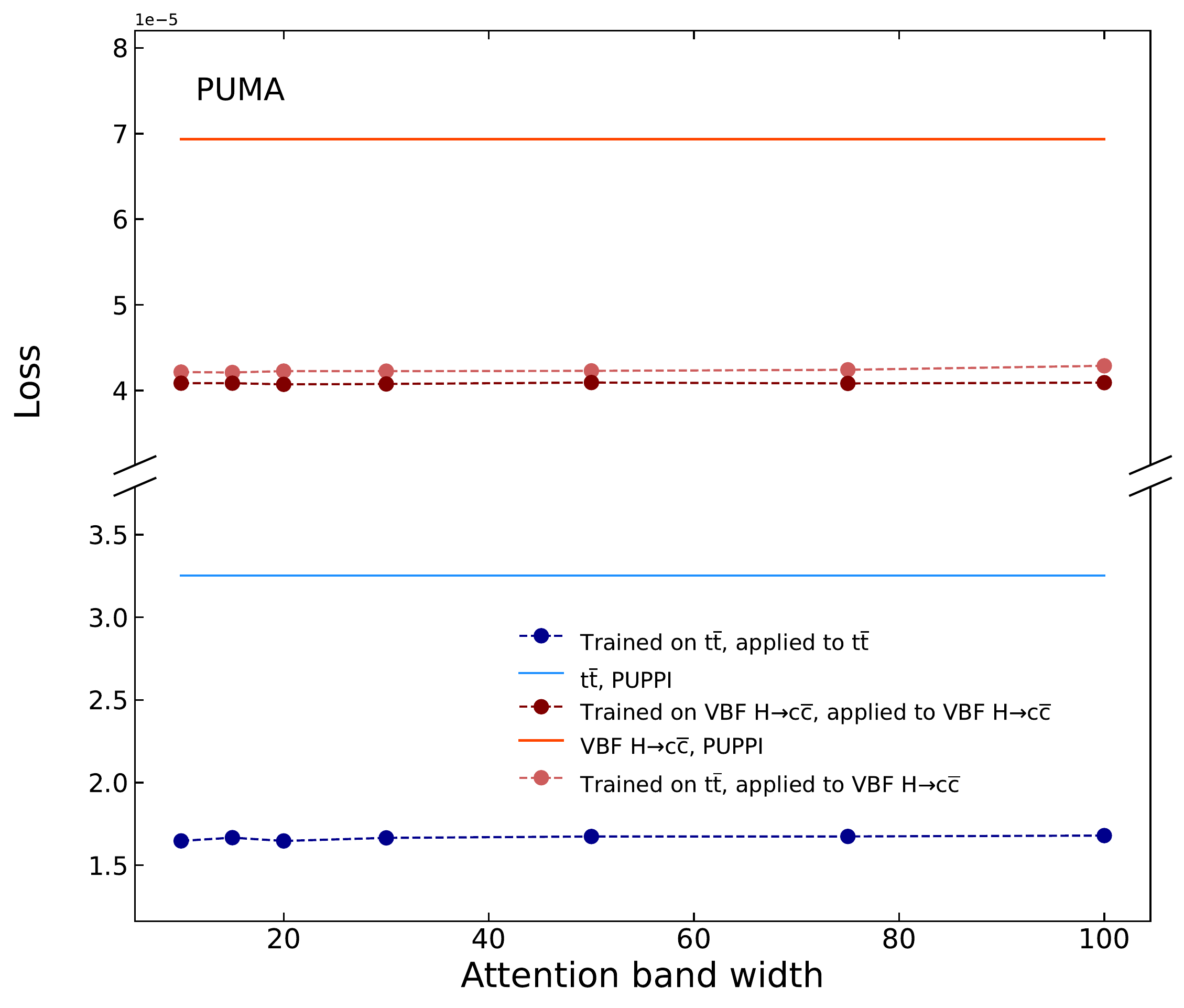}
  \includegraphics[width=0.455\textwidth]{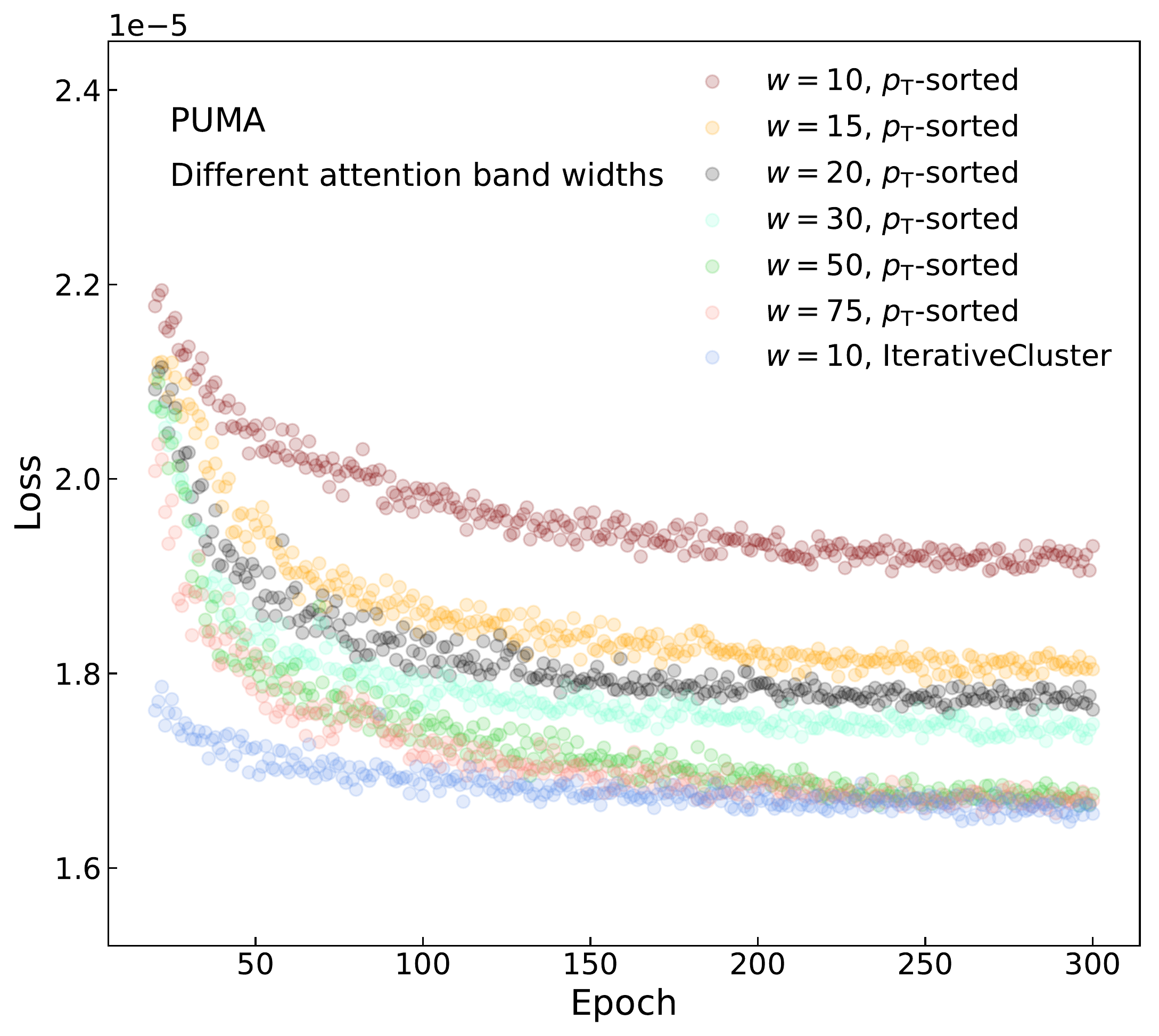}\\
  \caption{Left: \textsc{Puma} pile-up identification performance is not strongly sensitive to the attention band width, having fixed the cluster size at $w_\mathrm{clust}=10$. The blue points show the average loss on $\mathrm{t}\bar{\mathrm{t}}$ events, using the default \textsc{Puma} model trained on $\mathrm{t}\bar{\mathrm{t}}$ events. The light red points show the generalization of the default training to an entirely different process, VBF H($\mathrm{c}\bar{\mathrm{c}}$). Note that the difference in the absolute values of the two curves is not meaningful: the loss is dependent on a number of process-specific factors, like particle $p_\mathrm{T}$, $\eta$, charge, etc. The purpose of this figure is to demonstrate that in both processes, the loss is not strongly a function of $w$. The dark red points show the optimal result for VBF H($\mathrm{c}\bar{\mathrm{c}}$), i.e., when \textsc{Puma} is trained on this process instead of $\mathrm{t}\bar{\mathrm{t}}$. The increase in performance for this configuration is minimal considering the large improvement over PUPPI that is present for all \textsc{Puma} scenarios. Right: If the PF candidate sequence is ordered by $p_\mathrm{T}$, only very large attention band widths can recover the performance of \textsc{IterativeCluster} with an attention band width of 10.}
  \label{fig:att_width}
\end{figure}

Of particular interest is the attention band width, as this is chosen to be small to limit the memory footprint of the model. As shown in Fig.~\ref{fig:att_width}, it turns out \textsc{Puma} is robust across a range of attention widths. This holds true both for the process \textsc{Puma} has been trained on, as well as for the inference for a different physics process, namely Higgs boson production with subsequent decays of the Higgs boson to charm quarks. We hypothesize this is due to the optimal cluster ordering, describing pile-up primarily as a local problem; additionally, by stacking many Transformer layers and thereby increasing the receptive field, we are sensitive to residual long-range relations between pile-up particles even with small attention band widths. As noted in Sec.~\ref{subsec:band}, we do not explore $w>100$ due to computational constraints. 

\begin{figure}
  \centering
    \begin{subfigure}{0.42\textwidth} \includegraphics[width=\textwidth]{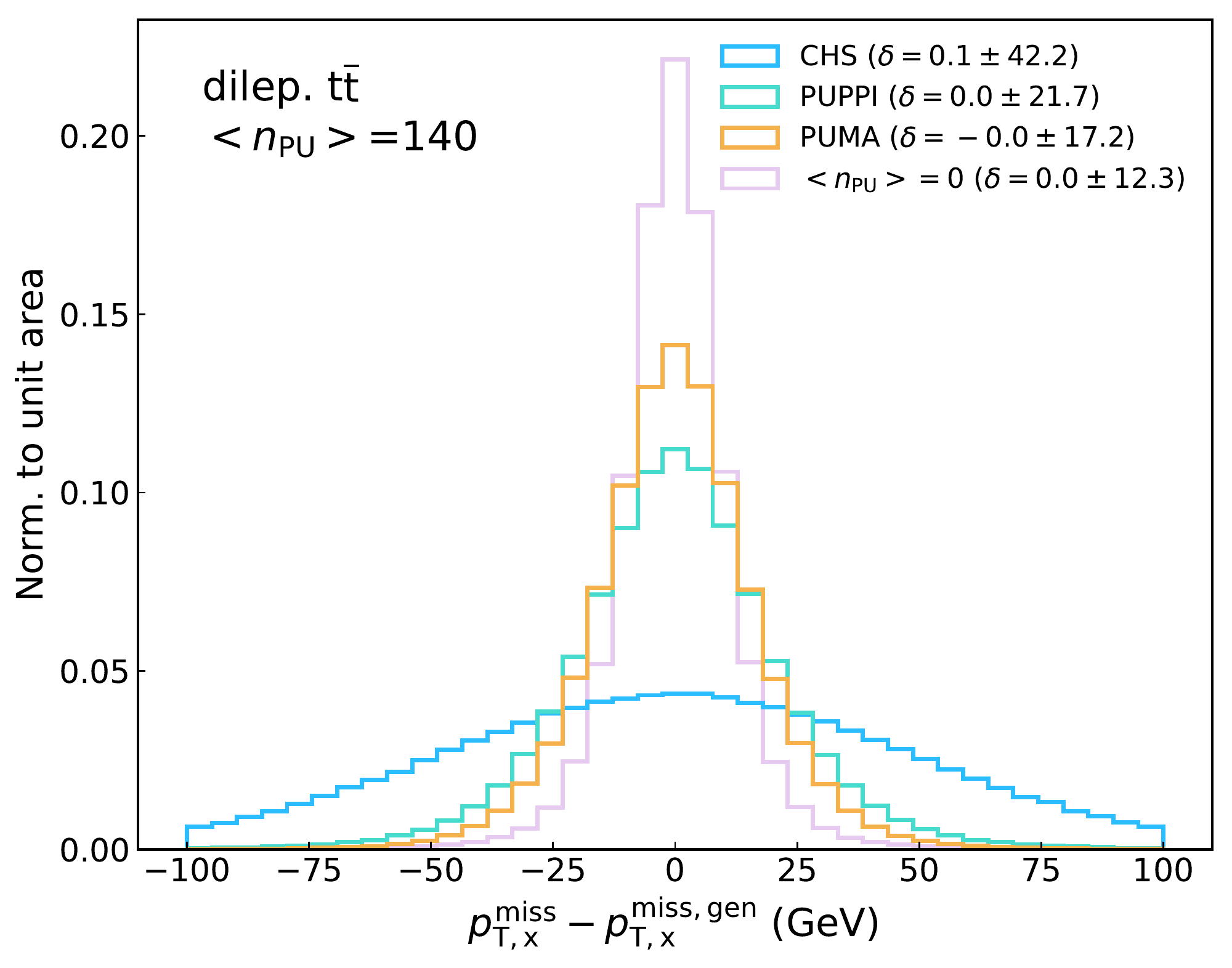} \caption{} \label{fig:metresx_ttbar} \end{subfigure}
  \begin{subfigure}{0.42\textwidth} \includegraphics[width=\textwidth]{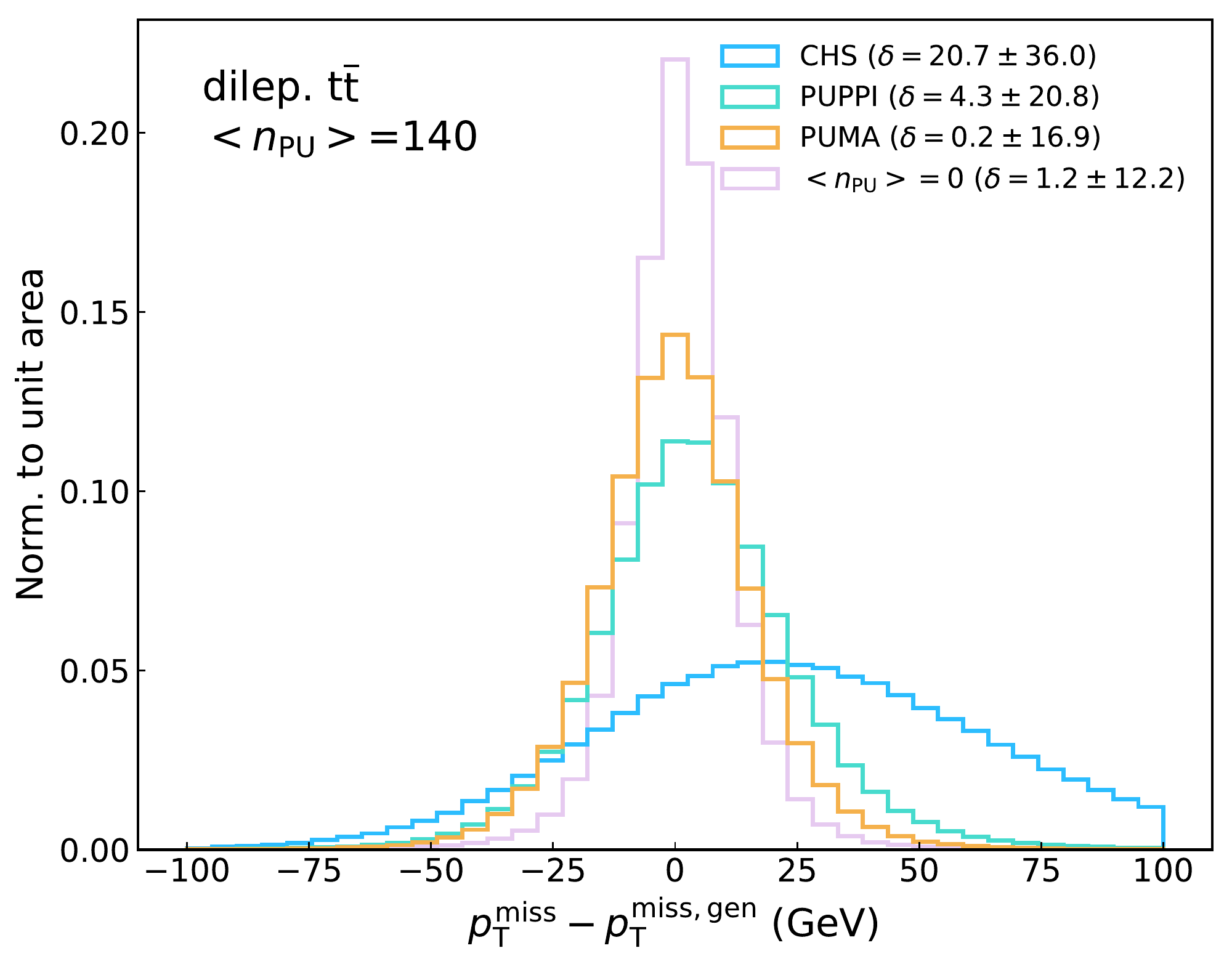} \caption{} \label{fig:metres_ttbar} \end{subfigure}\\
  \begin{subfigure}{0.42\textwidth} \includegraphics[width=\textwidth]{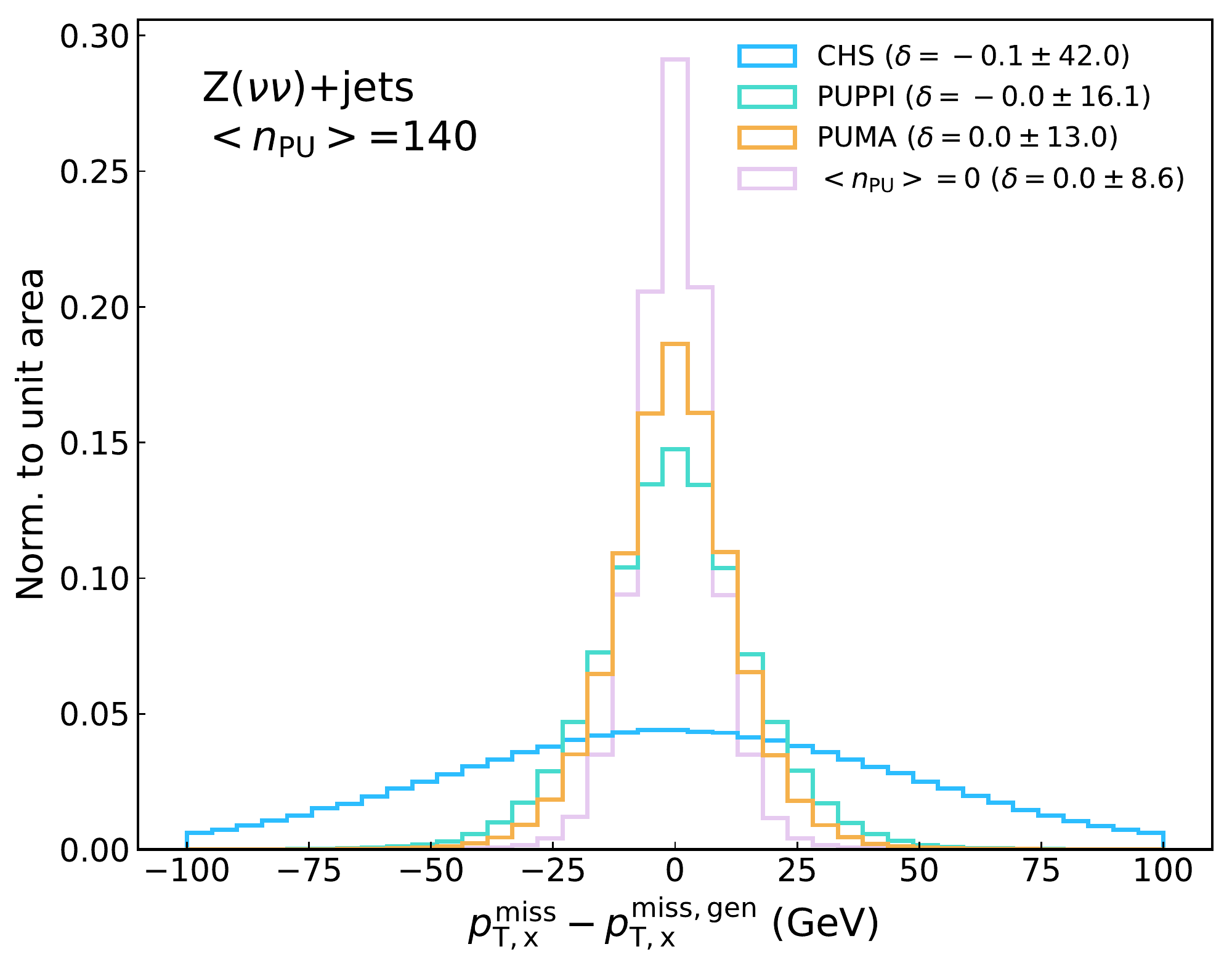} \caption{} \label{fig:metresx_zvvjets} \end{subfigure}
  \begin{subfigure}{0.42\textwidth} \includegraphics[width=\textwidth]{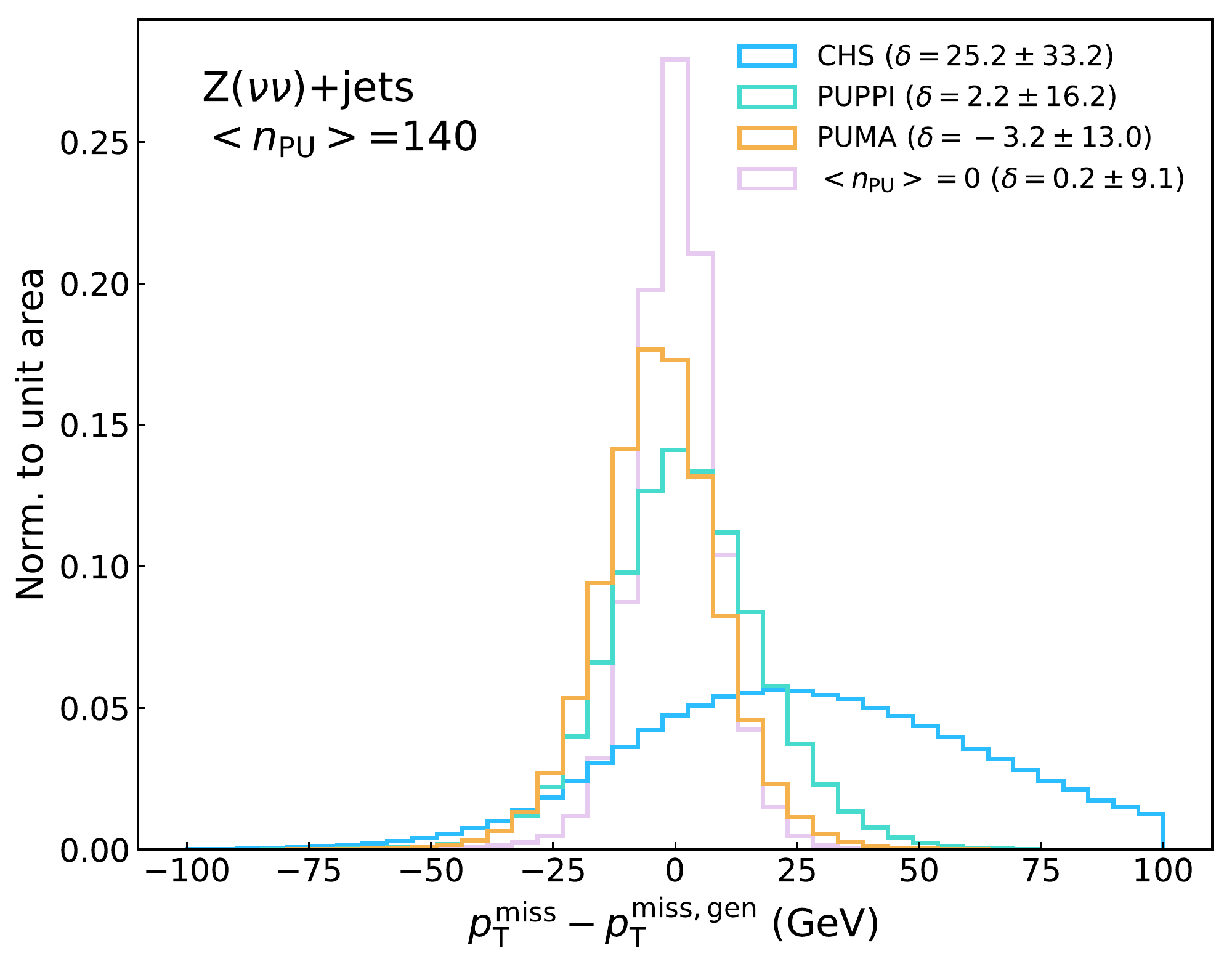} \caption{} \label{fig:metres_zvvjets} \end{subfigure}\\
    \begin{subfigure}{0.42\textwidth} \includegraphics[width=\textwidth]{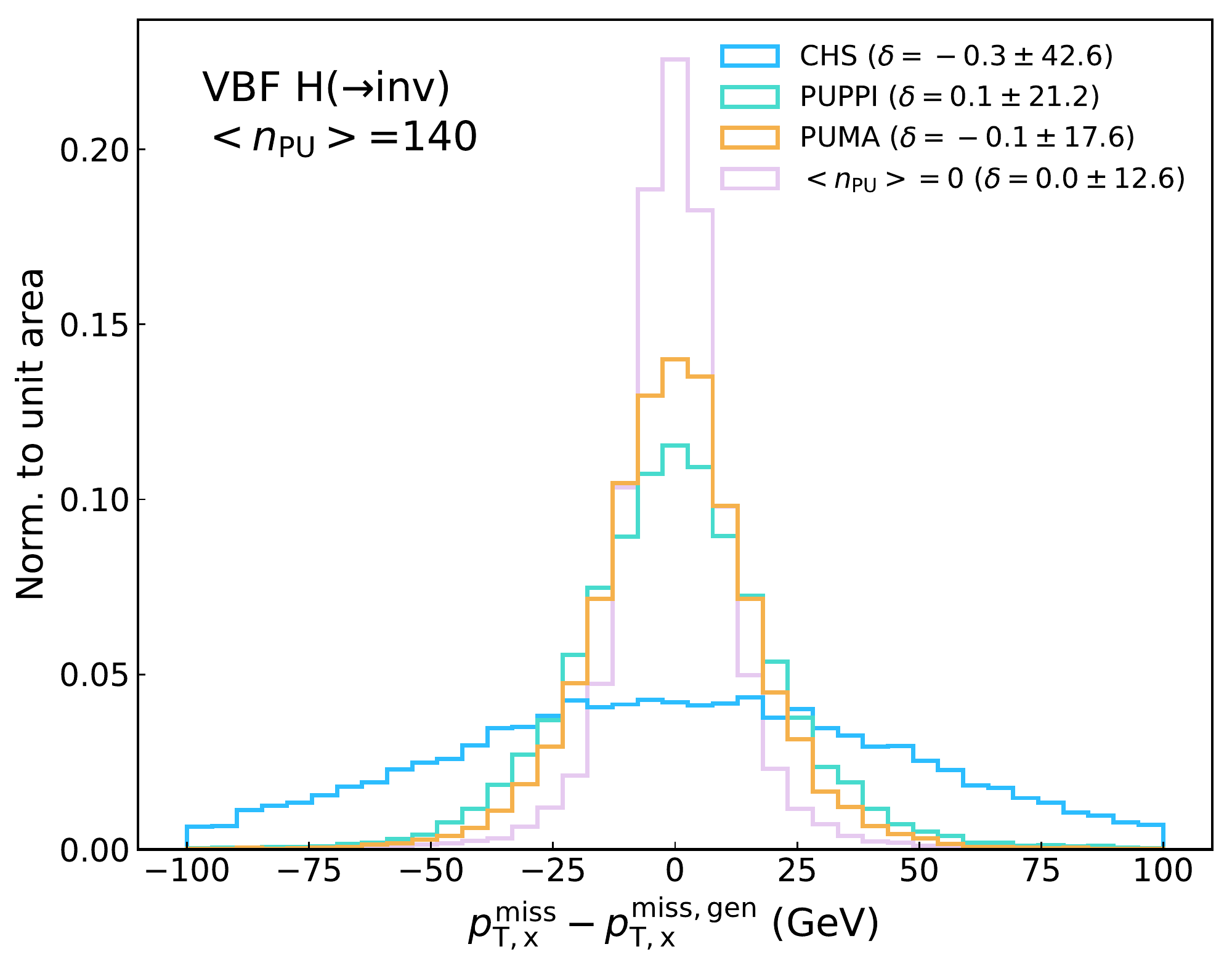} \caption{} \label{fig:metres_hinv} \end{subfigure}
    \begin{subfigure}{0.42\textwidth} \includegraphics[width=\textwidth]{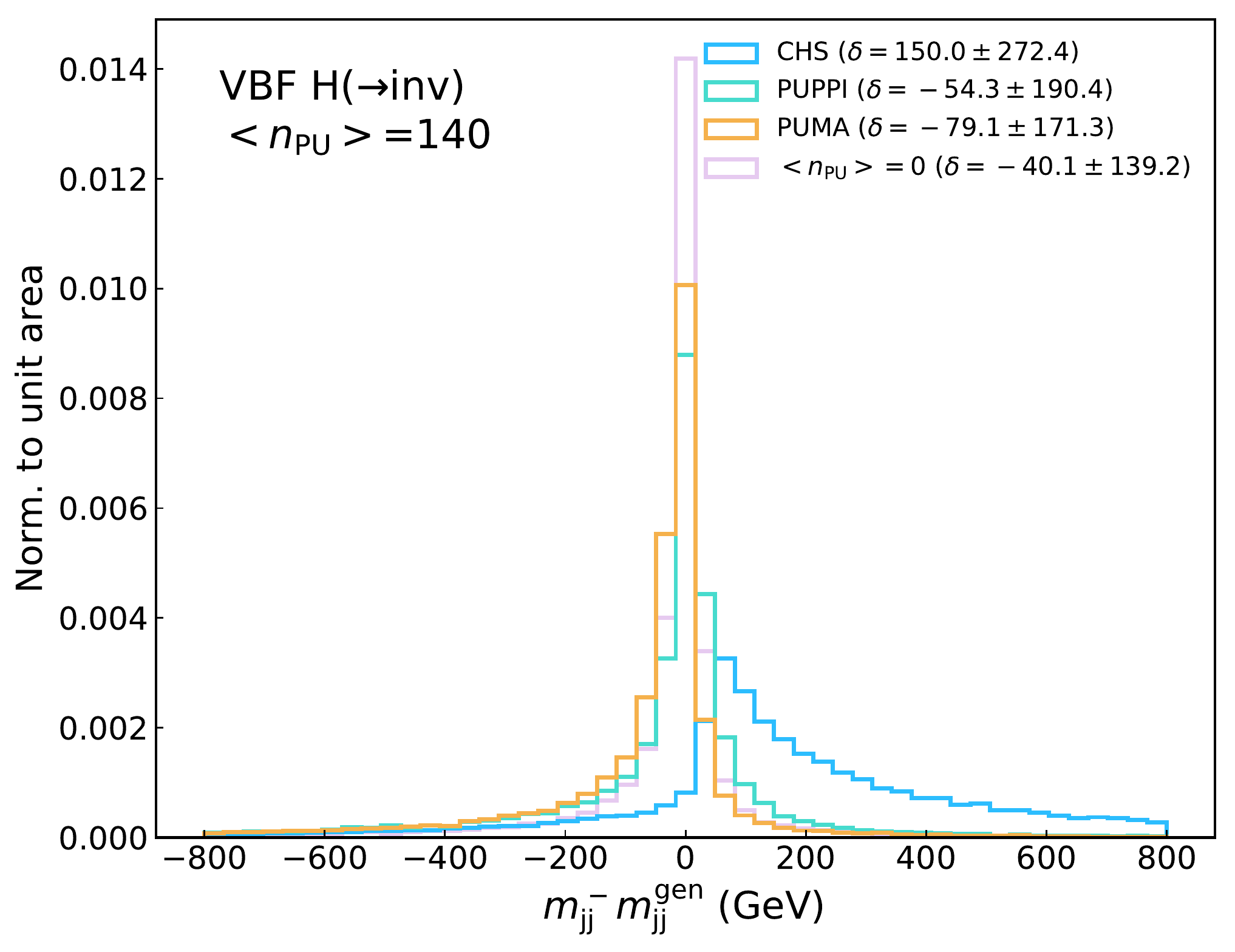} \caption{} \label{fig:jetres_hinv} \end{subfigure}\\

  \caption{Resolutions in $p_\mathrm{T}^\mathrm{miss}$ for top quark pair production with leptonic W boson decays in Figs.~\ref{fig:metresx_ttbar}-\ref{fig:metres_ttbar} are significantly improved for PUMA compared to other pile-up mitigation algorithms like CHS or PUPPI. Figures~\ref{fig:metresx_zvvjets}-\ref{fig:metres_zvvjets} show the resolution in $p_\mathrm{T}^\mathrm{miss}$ in Z($\nu\nu$)+jets events, i.e., for a different physics process from what \textsc{Puma} was trained on.  Figures \ref{fig:metres_hinv}-\ref{fig:jetres_hinv} show the resolution in $p_\mathrm{T}^\mathrm{miss}$ and, respectively, dijet mass in VBF Higgs production events, with the Higgs boson decaying to invisible dark matter particles. The parameter $\delta$ is estimated for each distribution and consists of the mean and the variance, reflecting the resolution in the estimation of the respective quantity.}
  \label{fig:res}
\end{figure}

\begin{figure}
  \centering
   \includegraphics[width=0.475\textwidth]{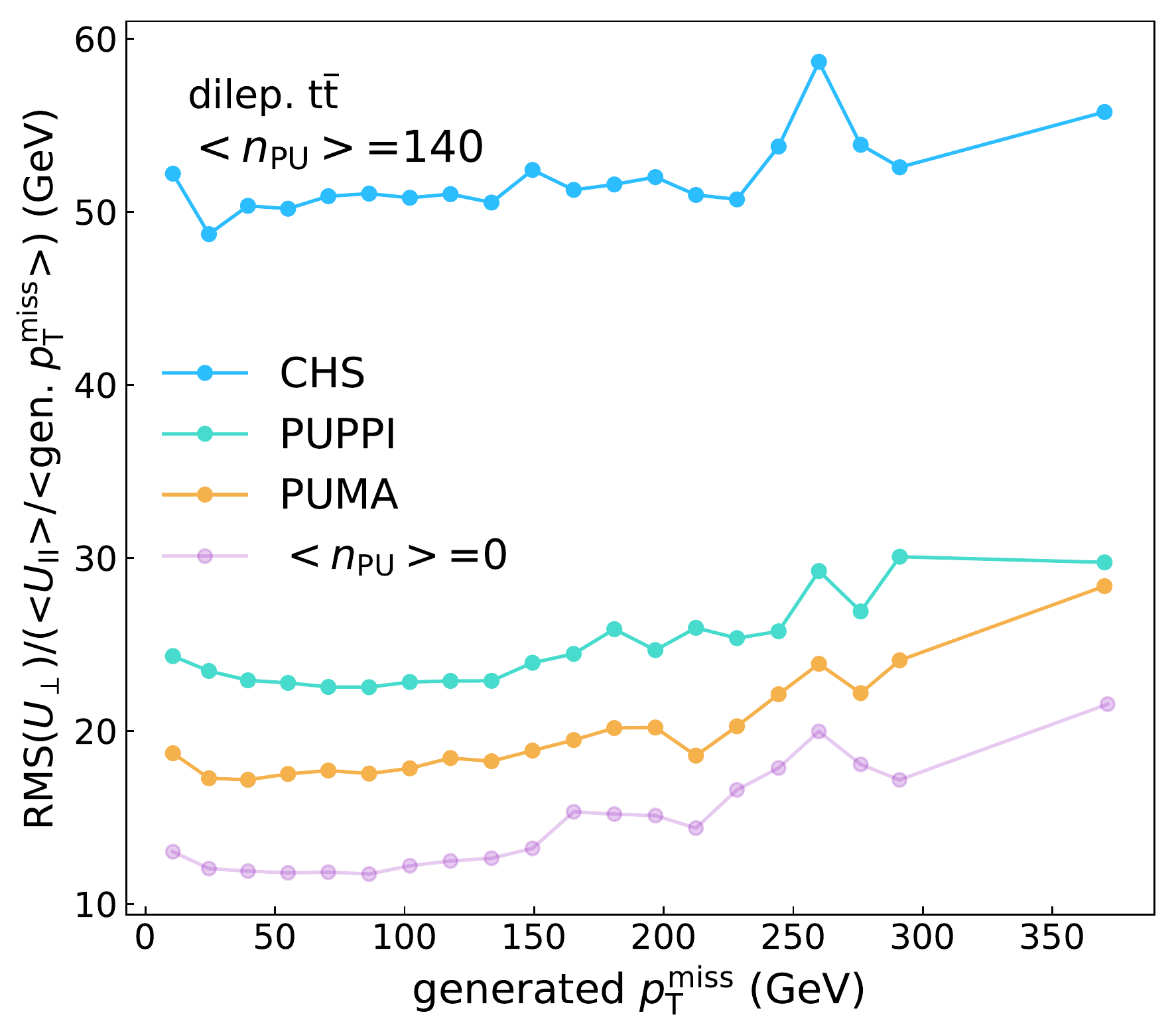}
 \includegraphics[width=0.475\textwidth]{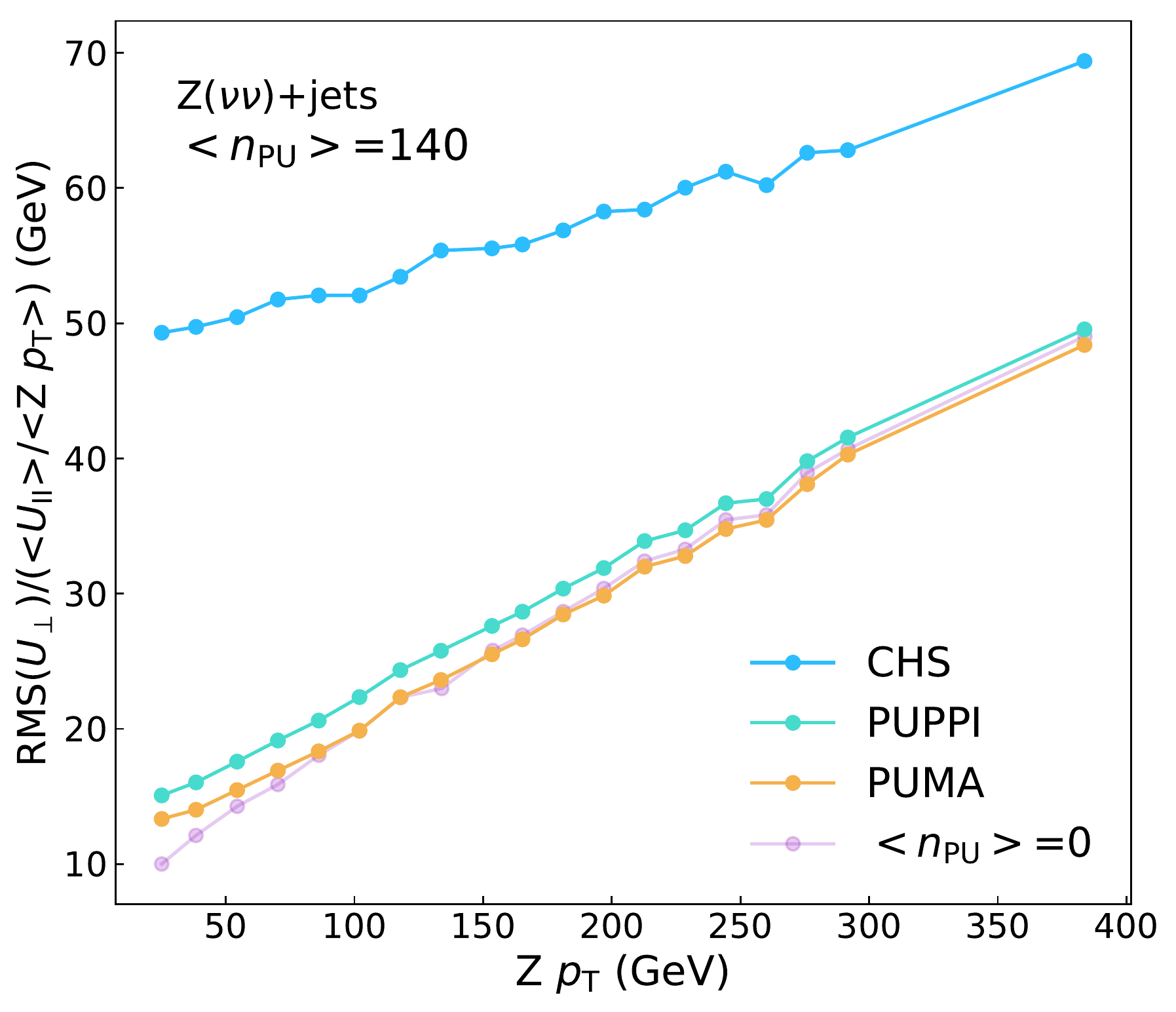}\\
  \caption{The RMS of the perpendicular hadronic recoil component for dileptonic top quark pair production and Z+jets production, as a function of $p_\mathrm{T}^\mathrm{miss,gen}$ and, respectively, the generated Z boson $p_\mathrm{T}$. The recoil is scaled to have unity response.}
  \label{fig:recoil}
\end{figure}

For inference, we consequently choose the model trained for 300 epochs with an attention band width of 15.

\subsection{Physics performance}

In terms of the key metrics introduced in Sec.~\ref{subsec:metrics}, we find that \textsc{Puma} outperforms PUPPI across the board. Figures~\ref{fig:metresx_ttbar}-\ref{fig:metres_ttbar} show the difference between estimated $p_\mathrm{T}^\mathrm{miss}$~and true $p_\mathrm{T}^\mathrm{miss}$~for the same process that \textsc{Puma} was trained on, i.e., dileptonic $\mathrm{t}\bar{\mathrm{t}}$ production. A sample that is statistically independent from the training sample has been used for inference. Note that the gold standard ($n_\mathrm{PU}=0$) does not have perfect reconstruction, due to finite detector and PF resolution. An absolute improvement in the $p_\mathrm{T}^\mathrm{miss}$~resolution, computed as the variance of the respective distribution, of about 20\% is observed compared to PUPPI. Alternatively, this can be characterized as \textsc{Puma} bridging half of the gap between PUPPI and the theoretical minimum gold standard.

% The model is able to bridge half of the gap between the resolution observed with PUPPI and the one observed in a sample without pile-up interactions. 

This demonstrates for the first time that a machine learning algorithm can mitigate pile-up effects more effectively than the currently best rule-based algorithm (PUPPI) at the event level, with a realistic detector simulation, and without actually using the scores of traditional algorithms like PUPPI as input features.

The same improvement is achieved for a different physics process in Figs.~\ref{fig:metresx_zvvjets}-\ref{fig:metres_zvvjets}, where the model was applied on a sample of Z($\nu\nu$)+jets events. Finally, as can be seen from Figs.~Figs.~\ref{fig:metres_hinv}-\ref{fig:jetres_hinv}, the dijet mass and $p_\mathrm{T}^\mathrm{miss}$ for VBF Higgs production events with invisible decays of the Higgs boson -- an essential search channel when looking for dark matter at the LHC -- likewise demonstrate an increase in resolution compared to PUPPI.

\begin{figure}
  \centering
   \includegraphics[width=0.475\textwidth]{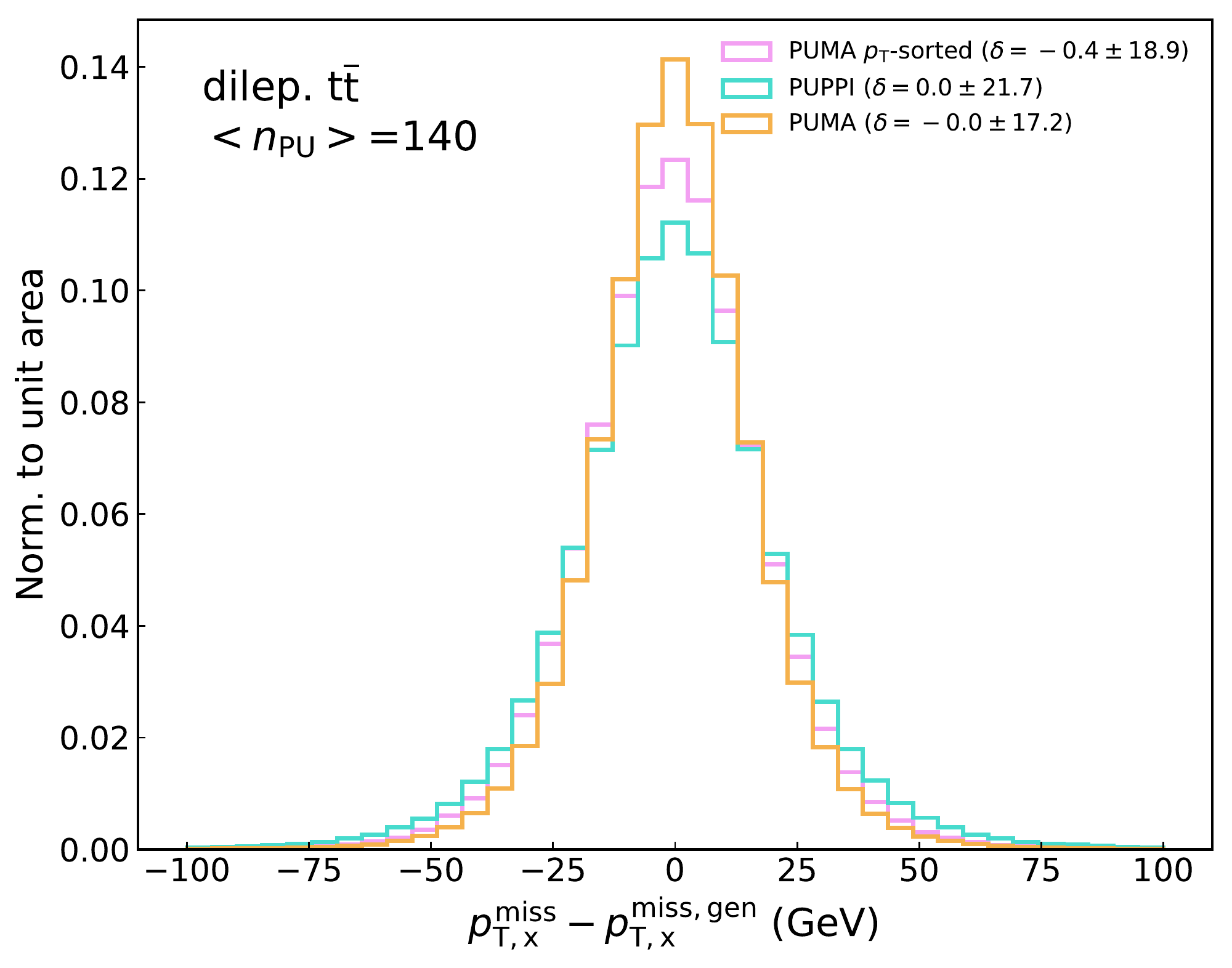}
   \includegraphics[width=0.475\textwidth]{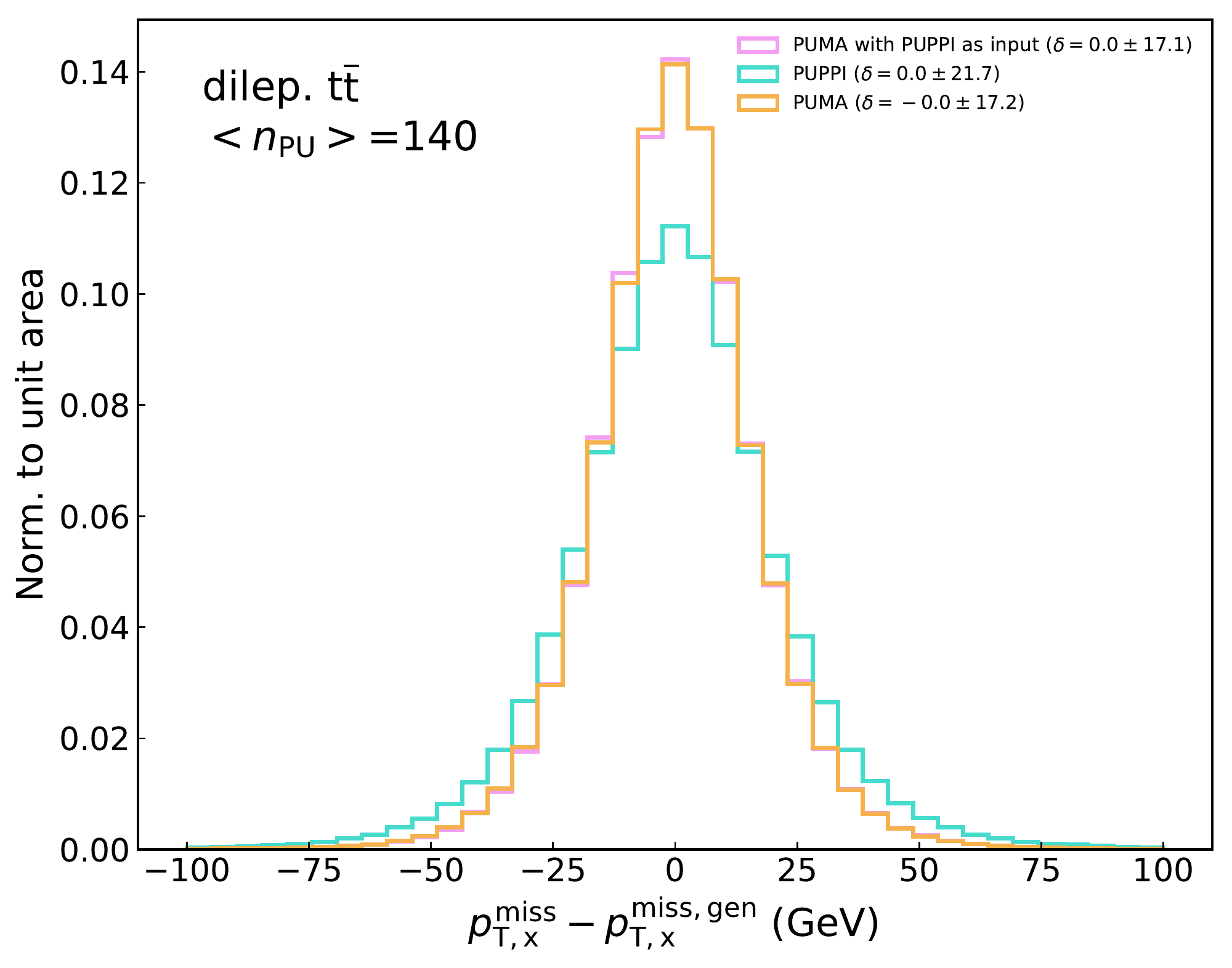}\\
  \caption{Left: Comparison between two \textsc{Puma} algorithms and PUPPI. An attention band width of 15 was used for both \textsc{Puma} models; the PF candidate sequence is either ordered by $p_\mathrm{T}$ or according to \textsc{IterativeCluster}, which gives the best performance. Right: Using the PUPPI weight per particle as input in addition to the variables listed in Tab.~\ref{tab:inputvars} does not result in sizable improvement over the default \textsc{Puma}. In both figures, the parameter $\delta$ is estimated for each distribution and consists of the mean and the variance, reflecting the resolution in the estimation of the respective quantity.}
  \label{fig:ptordering}
\end{figure}

The RMS error of the response-corrected $U_{\perp}$ distributions in $\mathrm{t}\bar{\mathrm{t}}$ production and Z+jets production is shown in Fig.~\ref{fig:recoil}, where a clear improvement over PUPPI is evident across the entire range of hadronic recoil.  By eliminating virtually all contributions from pile-up in the case of Z+jets, we obtain the same resolution as observed in the gold standard sample.

Finally, Fig.~\ref{fig:ptordering} demonstrates that the particle grouping found by \textsc{IterativeCluster} is necessary to achieve the observed performance. We compare to a version of \textsc{Puma} trained using $p_\mathrm{T}$-ordered particles. While momentum ordering does still lead to an improvement over PUPPI, this improvement is only half of what is observed with our optimal \textsc{Puma}. This demonstrates that pile-up is to first order a local problem, where the particles in the vicinity of the query particle contain the most information about its vertex of origin. It also shows that attention mechanisms combined with \textsc{IterativeCluster} are optimally suited to exploit this information. The marginal improvement shown in the right panel of Fig.~\ref{fig:ptordering} when using the PUPPI weight per particle as input feature in addition to the ones listed in Tab.~\ref{tab:inputvars} suggests that \textsc{Puma} manages to capture the information contained in PUPPI and further improves on it.

The observed improvement in the resolutions in $p_\mathrm{T}^\mathrm{miss}$ for a variety of processes, which are crucial for SM precision measurements and essential backgrounds in many searches for BSM physics, would directly translate to a more powerful analysis of a plethora of signatures expected in LHC collisions, especially towards the HL-LHC.

\section{Summary}
\label{sec:summary}

We have presented a highly effective pile-up mitigation algorithm \textsc{Puma} based on sparse self-attention. This method is the first machine learning approach relying only on raw reconstructed observables to demonstrate superior performance in a realistic detector scenario over current state-of-the-art, rule-based pile-up mitigation techniques. This holds true for both event-level and particle-level quantities. As pile-up effects will continue to worsen as the LHC moves to increasing luminosity, this is an important step towards showing that statistically-learned algorithms like sparse transformers can be very useful at the HL-LHC and beyond.

\section*{Acknowledgments}

The authors would like to thank Philip Harris and Lindsey Gray for productive conversations and feedback. The networks presented in this paper have been trained on the MIT-IBM Satori GPU cluster and on the Tier-2 MIT computing cluster. 

This material is based upon work supported by the U.S. National Science Foundation under Award Number PHY-1624356 and the U.S. Department of Energy Office of Science Office of Nuclear Physics under Award Number DE-SC0011939.

Disclaimer: ``This report was prepared as an account of work sponsored by an agency of the United
States Government. Neither the United States Government nor any agency thereof, nor any of their
employees, makes any warranty, express or implied, or assumes any legal liability or responsibility
for the accuracy, completeness, or usefulness of any information, apparatus, product, or process
disclosed, or represents that its use would not infringe privately owned rights. Reference herein to
any specific commercial product, process, or service by trade name, trademark, manufacturer, or
otherwise does not necessarily constitute or imply its endorsement, recommendation, or favoring by
the United States Government or any agency thereof. The views and opinions of authors expressed
herein do not necessarily state or reflect those of the United States Government or any agency thereof.''

\clearpage

\bibliography{puma}

%%end novalidate

\end{document}